\documentclass[journal]{IEEEtran}

\usepackage[utf8]{inputenc}
\usepackage[T1]{fontenc}
\usepackage{amsmath,amssymb,mathtools}
\usepackage{graphicx}
\usepackage{xcolor}
\usepackage{cite}
\usepackage{booktabs,array,multirow,tabularx}
\usepackage{setspace}
\usepackage{algorithm,algorithmic}
\usepackage{float}
\usepackage{subfig}
\usepackage{enumitem}
\usepackage{url}
\usepackage[nolist]{acronym}
\usepackage[input-decimal-markers=.]{siunitx}
\usepackage{pifont}
\usepackage{titlesec}
\usepackage{caption}
\usepackage{algorithm,algorithmic}
\usepackage{booktabs}
\usepackage[table]{xcolor}
\usepackage{array}
\usepackage{multirow}
\usepackage{siunitx}

\pdfoutput=1
\usepackage{amsthm}
\newtheorem{remark}{Remark}

\newcommand{\vect}[1]{\mathbf{#1}}
\newcommand{\norm}[1]{\left\lVert #1 \right\rVert_2}

\begin{acronym}
\acro{ACO}{asymmetrically clipped optical}
\acro{AWGN}{additive white Gaussian noise}
\acro{BER}{bit error rate}
\acro{CFR}{channel frequency response}
\acro{CIR}{channel impulse response}
\acro{CP}{cyclic prefix}
\acro{DC}{direct current}
\acro{FoV}{field of view}
\acro{FFT}{fast Fourier transform}
\acro{IFFT}{inverse fast Fourier transform}
\acro{IM}{index modulation}
\acro{IM/DD}{intensity modulation/direct detection}
\acro{ICL}{integrated communication and localisation}
\acro{LED}{light-emitting diode}
\acro{LoS}{line-of-sight}
\acro{LS}{least squares}
\acro{NLoS}{non-LoS}
\acro{OFDM}{orthogonal frequency-division multiplexing}
\acro{OWS}{optical wireless system}
\acro{PD}{photodetector}
\acro{QAM}{quadrature amplitude modulation}
\acro{RIS}{reconfigurable intelligent surface}
\acro{RMSE}{root mean-square error}
\acro{RSS}{received signal strength}
\acro{SA}{sensing-assisted}
\acro{SNR}{signal-to-noise ratio}
\acro{DFT}{discrete Fourier transform}
\acro{CE}{channel estimation}
\acro{DCO}{direct-current offset}
\acro{SE}{spectral efficiency}
\acro{NMSE}{normalised mean squared error}
\acro{CFMO}{clipping-free multilayer optical}
\acro{ACO}{asymmetrically clipped optical}
\acro{LACO}{layered ACO}
\acro{RF}{radio-frequency}
\acro{PE}{power-efficiency}
\acro{FD}{frequency-domain}
\acro{TD}{time-domain}
\acro{PF}{positive-frequency}
\acro{ML}{maximum likelihood}
\end{acronym}

\titlespacing{\subsection}{1pt}{*0.5}{*0.35}
\titlespacing{\subsubsection}{3pt}{*0.45}{*0.25}
\begin{document}

\title{\LARGE Sensing-Assisted OFDM with Index Modulation: A Waveform for Optical Wireless Integrated Communication and Localisation}

\author{Ali Waqar Azim, Mohamed Sufyan Islim, Muhammad Ali Imran, Harald Hass, and Hanaa Abumarshoud%
\thanks{Ali Waqar Azim, Muhammad Ali Imran, and Hanaa Abumarshoud are with the James Watt School of Engineering, University of Glasgow, Glasgow G12 8QQ, U.K.
(email: \{ali.azim, muhammad.imran, hanaa.abumarshoud\}@glasgow.ac.uk).\\Mohamed Sufyan Islim is with pureLiFi Ltd., Edinburgh EH6 6QH, U.K. (email: mohamed.islim@purelifi.com). \\Harald Haas is with Department of Engineering, University of Cambridge,
9 JJ Thomson Avenue, Cambridge CB3 0FA, UK (e-mail: huh21@cam.ac.uk).}}
\maketitle
\begin{abstract}
This paper proposes sensing-assisted OFDM with index modulation (SA-OFDM-IM) for joint communication and localisation over optical wireless systems. In SA-OFDM-IM, a \textit{multilayer frequency-domain architecture} partitions each layer's admissible subcarriers into disjoint \textit{communication} and \textit{distributed sensing-pilot} subsets, applying group-wise index modulation (IM) within the communication subset. 
A constrained sensing-to-communication \textit{energy allocation} jointly minimises average bit-error rate (BER) and delay-based localisation root mean square error (RMSE) under a minimum power-domain margin, enabling \textit{low-complexity}, ordered-energy detection of both sensing and communication subcarriers without exhaustive maximum-likelihood search. Sensing tones are aggregated across the frame to reconstruct the channel frequency response, yielding the line-of-sight (LoS) delay tap and phase-compensated LoS gain for multilateration and, via the proposed \textit{delay-aided received signal strength (RSS) refinement}. Simulation results indicate that SA-OFDM-IM can achieve up to \(5\) dB lower SNR requirement in BER and up to three orders of magnitude lower channel-estimation error than DC-biased optical OFDM, while the RSS estimator removes the SNR-independent error floors of delay-only and hybrid-RSS benchmarks to attain sub-delay-bin range and position accuracy, remaining robust under \textit{random known and unknown receiver orientations}.
\end{abstract}
\vspace{-2mm}
\IEEEpeerreviewmaketitle
\vspace{-6mm}
\section{Introduction}
\label{sec:introduction}
\Ac{ICL} has emerged as a foundational paradigm for future wireless networks, enabling communication and sensing to be co-designed over shared spectral, temporal, and hardware resources. While \ac{ICL} has reached relative maturity in \ac{RF} systems, its extension to \acp{OWS} remains comparatively nascent. Nevertheless, \ac{OWS} is a naturally favourable platform for integrated operation as: it exploits a vast, licence-free optical spectrum, is inherently immune to \ac{RF} electromagnetic interference, offers strong spatial confinement, and can reuse existing illumination infrastructure with \textit{a priori} known \ac{LED} luminaire geometry  for both communication and sensing \cite{wen2024opticalISAC,shi2024experimentalISAC}.

From a waveform-design standpoint, optical \ac{OFDM} is particularly well suited to \ac{OWS}, as it permits single-tap equalisation and, in an \ac{ICL} setting, affords considerable flexibility in allocating subcarriers to communication or sensing functions. Because \acp{PD} respond only to instantaneous optical intensity, \ac{IM/DD}-compatible \ac{OFDM} variants must produce a real-valued, non-negative \ac{TD} signal, enforced through Hermitian-symmetry with biasing or clipping, as in \ac{DCO}-\ac{OFDM} and \ac{ACO}-\ac{OFDM} \cite{dissanayake2013comparison,wang2023clipping}. While \ac{DCO}-\ac{OFDM} generally achieves higher \ac{SE}, its required bias imposes a \ac{PE} penalty. This trade-off has motivated power-efficient layered alternatives such as \ac{LACO}-\ac{OFDM} \cite{wang2015layered} and \ac{CFMO}-\ac{OFDM} \cite{wang2023clipping}, which superimpose multiple \ac{FD} layers to approach the \ac{SE} of \ac{DCO}-\ac{OFDM}. However, \ac{LACO}-\ac{OFDM} incurs considerable receiver decoding complexity, which \ac{CFMO}-\ac{OFDM} alleviates while preserving favourable \ac{SE}-\ac{PE} characteristics \cite{wang2023clipping}, yet extending these waveforms to optical \ac{ICL} remains largely unexplored, a gap Xu \textit{et al.} \cite{xu2025integrated} recently addressed in part with a \ac{DCO}-\ac{OFDM}-based optical \ac{ICL} framework demonstrating joint communication and localisation.

More recently, \ac{OFDM} with \ac{IM} has been introduced for \acp{OWS}, wherein part of the bit stream is conveyed through the combinatorial activation pattern that switches subcarriers on within each group, besides constellation symbols \cite{basar2015opticalIM}. Since the number of admissible activation patterns for a group of \(g\) subcarriers with \(k\) active tones grows combinatorially as \(\binom{g}{k}\), the index domain constitutes an additional information-bearing dimension yielding \(\lfloor\log_2\binom{g}{k}\rfloor\) extra bits per group, thereby offering a flexible \ac{SE}-\ac{PE} trade-off relative to conventional \ac{OFDM} \cite{azim2020laco,azim2021filtered}. Although \ac{OFDM}-\ac{IM} has already been investigated for \ac{RF} \ac{ICL} systems \cite{elbir2024index}, its extension to optical \ac{ICL} is non-trivial because an optical \ac{OFDM}-\ac{IM} waveform must simultaneously: (i) maintain disjoint, unambiguously separable communication and sensing subcarriers within the same frame, (ii) supply sufficiently dense \ac{FD} pilot observations for reliable \ac{CE} and delay/gain-based localisation, and (iii) admit low-complexity receiver processing that avoids exhaustive joint \ac{ML} search over the combined index-pattern and constellation-symbol hypothesis space.

Indoor \acp{OWS} typically employ ceiling-mounted \acp{LED} at known coordinates as transmitters and \acp{PD} as receivers. Beyond their communication role, the known \ac{LED} coordinates, the inverse-square distance-dependent optical path loss, and the direction-dependent Lambertian radiation/incidence pattern jointly furnish geometrically informative sensing observables for localisation. Optical positioning can accordingly be realised via propagation-delay ranging, \ac{RSS}-based ranging, or hybrid combinations thereof. Delay-based ranging is attractive because the extracted \ac{LoS} tap is largely insensitive to \ac{PD} orientation as long as the \ac{PD}'s \ac{FoV} is not blocked, and does not require accurate optical-gain calibration. However, its accuracy is fundamentally bounded by the underlying discrete-time channel estimate and by the available bandwidth \(B\), since the associated ranging quantisation step scales as \(\frac{c_0}{B}\), where \(c_0\) is the speed of light. In contrast, \ac{RSS}-based localisation can, in principle, achieve sub-delay bin range accuracy by exploiting variations in received optical power, but its reliability is comparatively more sensitive to \ac{PD} orientation, \ac{FoV}, and unresolved multipath components \cite{xu2025integrated}. Furthermore, existing optical localisation frameworks built on optical \ac{OFDM} typically rely on dedicated training symbols or separately reserved sensing resources for \ac{CE} \cite{xu2025integrated,shi2024experimentalISAC}. However, reserving an entire \ac{OFDM} symbol within a frame for \ac{CE} negatively impacts the achievable \ac{SE}. Another important parameter that impacts the localisation accuracy is the \ac{PD}'s orientation in mobile \ac{OWS} deployments, since the incidence-angle projection directly scales the received \ac{LoS} gain. Thus, \ac{RSS}-based refinement reliability degrades markedly when the \ac{PD} orientation is unknown. Measurement-based studies indicate that handheld-device orientations are inherently random and are more accurately characterised by a statistical polar-azimuth model, e.g., a Laplacian-distributed tilt angle and a uniformly distributed azimuth angle \cite{soltani2018modeling}. Consequently, a practical optical localisation architecture must explicitly accommodate both known and unknown \ac{PD} orientations, while preventing orientation-induced attenuation from being misattributed to a variation in propagation distance.

Motivated by these observations, this paper proposes \ac{SA}-\ac{OFDM}-\ac{IM}, a waveform that jointly integrates index-domain communication and distributed sensing pilots within a multilayer \ac{FD} frame structure, drawing inspiration from \ac{CFMO}-\ac{OFDM} \cite{wang2023clipping}. Within each layer, the admissible subcarriers are partitioned into communication and sensing subcarriers, and a constrained power-allocation strategy enforces a minimum sensing-to-communication energy separation enabling low-complexity, order-statistics-based subcarrier detection while jointly balancing average \ac{BER} and localisation \ac{RMSE}. At the \ac{PD}, the sensing-pilot observations are coherently aggregated across the frame to estimate the \ac{CFR}/\ac{CIR}, from which the \ac{LoS} delay tap and phase-compensated \ac{LoS} gain are extracted. These two sensing observables are subsequently exploited by a delay-aided \ac{RSS} refinement stage, which uses the coarse delay-based inputs to jointly estimate the \ac{PD} position and, when unknown, its surface-normal orientation vector.

To this end, an \ac{OFDM}-\ac{IM}-based waveform designed specifically in context of optical \ac{ICL} has not previously been reported. Against this background, the contributions of this paper are summarised as follows:
\vspace{-2mm}
\begin{itemize}[leftmargin=*]
    \item \ac{SA}-\ac{OFDM}-\ac{IM} waveform is proposed for optical \ac{ICL}, in which communication symbols and sensing pilots are embedded through structured, dyadic subcarrier allocation across successive \ac{FD} layers, while satisfying the \ac{IM/DD} non-negativity constraint via an adaptive periodic bias.
    \item A constrained sensing-communication energy-allocation framework is formulated as weighted bi-objective optimisation, jointly incorporating average \ac{BER} and average delay-based localisation \ac{RMSE} evaluated over a representative \ac{SNR} set, subject to a minimum power-domain separation that facilitates low-complexity sensing and communication subcarrier detectability.
    \item A receiver signal-processing chain is established that extracts both communication and sensing observables, integrating energy-ordered subcarrier detection, pilot-aided \ac{LS} \ac{CE}, frame-level coherent \ac{CFR} aggregation, and delay-domain interpolation to recover the channel and delay information, alongside joint recovery of index- and constellation-domain information for communication.
    \item A delay-aided \ac{RSS}-based localisation method is introduced, in which the \ac{LoS} delay supplies a multilateration-based initial position estimate that both initialises and safeguards the subsequent refinement, while the phase-compensated \ac{LoS} gains refine the \ac{PD} position and, when required, jointly estimate its surface-normal vector.
\end{itemize}
Numerical results demonstrate improved \ac{BER} and \ac{CE} accuracy for the proposed \ac{SA}-\ac{OFDM}-\ac{IM} relative to \ac{DCO}-\ac{OFDM}. By aggregating distributed sensing-pilot observations across the frame, the proposed scheme improves the estimated \ac{CFR}/\ac{CIR} and the extracted \ac{LoS} gains, enabling the delay-aided \ac{RSS} estimator to achieve substantially lower range and position \acp{RMSE} than delay-only and hybrid-\ac{RSS} benchmarks, under upward-facing, randomly oriented known, and randomly oriented unknown \ac{PD} configurations.

The remainder of this paper is organised as follows. Section~\ref{sec:optical_channel_model} presents the optical channel model, Section~\ref{sec2} explains the proposed waveform design and \ac{ICL} framework, and energy-allocation strategy, and Section~\ref{sec:delay_rss_ranging_position} the delay- and \ac{RSS}-based ranging, position, and orientation estimators. Section~\ref{sim_diss} presents numerical results, and Section~\ref{sec:conclusion} concludes the paper.
\vspace{-5mm}
\section{Optical Channel Model}
\label{sec:optical_channel_model}
Consider an indoor \ac{OWS} comprising \(J\) ceiling-mounted \acp{LED}, indexed by \(j\in\{1,2,\ldots,J\}\), each acting as a spatially distributed transmit anchor with known coordinates \(\vect{c}_{\mathrm S,j}=[x_{\mathrm S,j},y_{\mathrm S,j},z_{\mathrm S,j}]^{\mathrm T}\) in the global Cartesian reference frame, while the \ac{PD} is placed at \textit{a priori} unknown coordinates \(\vect{c}_{\mathrm D}=[x_{\mathrm D},y_{\mathrm D},z_{\mathrm D}]^{\mathrm T}\). The Euclidean separation between \ac{LED} \(j\) and the \ac{PD} is \(d_{0,j}=\norm{\vect{c}_{\mathrm D}-\vect{c}_{\mathrm S,j}}\), where \(\norm{\cdot}\) denotes the Euclidean norm, resulting in \ac{LoS} propagation delay of \(\tau_{0,j}=\frac{d_{0,j}}{c_0}\). The corresponding forward and reverse unit vectors are \(\vect{u}_{\mathrm{SD},j}=\frac{\vect{c}_{\mathrm D}-\vect{c}_{\mathrm S,j}}{d_{0,j}}\) and \(\vect{u}_{\mathrm{DS},j}=-\vect{u}_{\mathrm{SD},j}\), respectively, which parameterise the emergence and incidence angles that govern the directional radiation and collection efficiency.

The radiation directivity of \ac{LED} \(j\) is captured by its surface-normal vector \(\vect{n}_{\mathrm S,j}\). For a ceiling-mounted luminaire tilted by an angle \(\theta\) from the downward-facing boresight, the normal is expressed as \(\vect{n}_{\mathrm S,j}=[\sin(\pi-\theta),\sin(\pi-\theta),\cos(\pi-\theta)]^{\mathrm T}\) \cite{xu2025integrated}, which reduces to \(\vect{n}_{\mathrm S,j}=[0,0,-1]^{\mathrm T}\) in the untilted, strictly downward-facing case (\(\theta=0\)). Conversely, the \ac{PD} orientation can also be treated as random in line with empirically validated mobile-receiver behaviour \cite{soltani2018modeling}. Adopting the polar-azimuth representation of \cite{soltani2018modeling}, the \ac{PD} surface normal is \(\vect{n}_{\mathrm D}(\theta_{\mathrm D},\omega_{\mathrm D})=[\sin\theta_{\mathrm D}\cos\omega_{\mathrm D},\sin\theta_{\mathrm D}\sin\omega_{\mathrm D},\cos\theta_{\mathrm D}]^{\mathrm T}\), where the polar tilt angle \(\theta_{\mathrm D}\) follows Laplacian distribution, and the azimuth angle \(\omega_{\mathrm D}\) is uniformly distributed over \([-180^\circ,180^\circ]\).

The \ac{LoS} emergence and incidence angle cosines follow directly from the normal vectors as \(\cos\phi_{t,0,j}=\vect{u}_{\mathrm{SD},j}^{\mathrm T}\vect{n}_{\mathrm S,j}\) and \(\cos\phi_{r,0,j}=\vect{u}_{\mathrm{DS},j}^{\mathrm T}\vect{n}_{\mathrm D}\). At the \ac{PD}, a concentrator of refractive index \(n_{\mathrm c}\) restricts collection to incidence angles within the \ac{FoV} semi-angle \(\Psi_{\mathrm{FoV}}\), giving the concentrator gain \(g_{\mathrm{oc},j}(\phi_{r,0,j})=\frac{n_{\mathrm c}^{2}}{\sin^{2}\Psi_{\mathrm{FoV}}}\) for \(0\leq\phi_{r,0,j}\leq\Psi_{\mathrm{FoV}}\), and zero otherwise.
On the transmit side, the \ac{LED} radiation pattern is modelled as generalised Lambertian with order \(q=-\frac{\ln 2}{\ln(\cos\phi_{1/2})}\), where \(\phi_{1/2}\) denotes the \ac{LED} half-power semi-angle and governs the directional concentration of the emitted intensity.

Combining the emission, propagation, and collection stages yields the \ac{LoS} channel gain between \ac{LED} \(j\) and the \ac{PD}:
\begin{equation}
\alpha_{0,j}
=
\frac{(q+1)A_{\mathrm{PD}}}{2\pi d_{0,j}^{2}}
\cos^{q}(\phi_{t,0,j})
\cos(\phi_{r,0,j})
g_{\mathrm{of}}
g_{\mathrm{oc},j}(\phi_{r,0,j}),
\label{eq:los_gain}
\end{equation}
where \(A_{\mathrm{PD}}\) is the \ac{PD} active detection area and \(g_{\mathrm{of}}\) is the fixed insertion gain of the optical bandpass filter. The channel gain in \eqref{eq:los_gain} is zero whenever \(\cos\phi_{t,0,j}\leq0\), \(\cos\phi_{r,0,j}\leq0\), or \(\phi_{r,0,j}>\Psi_{\mathrm{FoV}}\), i.e., whenever the \ac{PD} lies outside the \ac{LED}'s forward radiating hemisphere or the incidence angle exceeds the concentrator's admissible \ac{FoV}.

Beyond the direct \ac{LoS} ray, the optical channel also has reflected \ac{NLoS} components from diffuse surface scattering. The \(\rho\)-th such path, \(\rho=1,\ldots,P_j-1\), has a propagation delay \(\tau_{\rho,j}\) and gain \(\alpha_{\rho,j}\), with \(P_j\) the number of resolvable \ac{LoS}/\ac{NLoS} paths for \ac{LED} \(j\), yielding a sparse tapped-delay-line \ac{CIR} as:
\begin{equation}
h_j(\tau)=\sum_{\rho=0}^{P_j-1}\alpha_{\rho,j}\delta_{\mathrm D}(\tau-\tau_{\rho,j}),
\end{equation}
where \(\delta_{\mathrm D}(\cdot)\) is the Dirac-delta function and \(\rho=0\) is reserved for the \ac{LoS} contribution of \eqref{eq:los_gain}.
The continuous-time \ac{CIR} is discretised onto a uniform sampling grid of interval \(T_{\mathrm{sam}}=\frac{1}{B}\), so each path delay \(\tau_{\rho,j}\) is quantised to the nearest sampling instant as the discrete delay tap \(\lambda_{\rho,j}=\frac{\tau_{\rho,j}}{T_{\mathrm{sam}}}=\tau_{\rho,j}B\), with \(\lfloor\cdot\rceil\) rounding to the nearest integer. This mapping ties the achievable delay resolution to the system bandwidth, since \(T_{\mathrm{sam}}\) sets the quantisation step of the delay grid, and yields the discrete-time \ac{CIR} as:
\begin{equation}
h_j[\lambda]=\sum_{\rho=0}^{P_j-1}\alpha_{\rho,j}\delta_\mathrm{K}[\lambda-\lambda_{\rho,j}]   
\end{equation}
for \(\lambda=0,1,\ldots,L_h-1\), where \(\delta_{\mathrm K}(\cdot)\) is the Kronecker delta and the effective channel length.
%
\vspace{-3mm}
\section{Proposed SA-OFDM-IM Waveform}\label{sec2}
\begin{figure*}[ht]
    \centering
    \includegraphics[
        width=0.97\textwidth,
        trim=0.2cm 0.2cm 0.18cm 0.2cm,
        clip
    ]{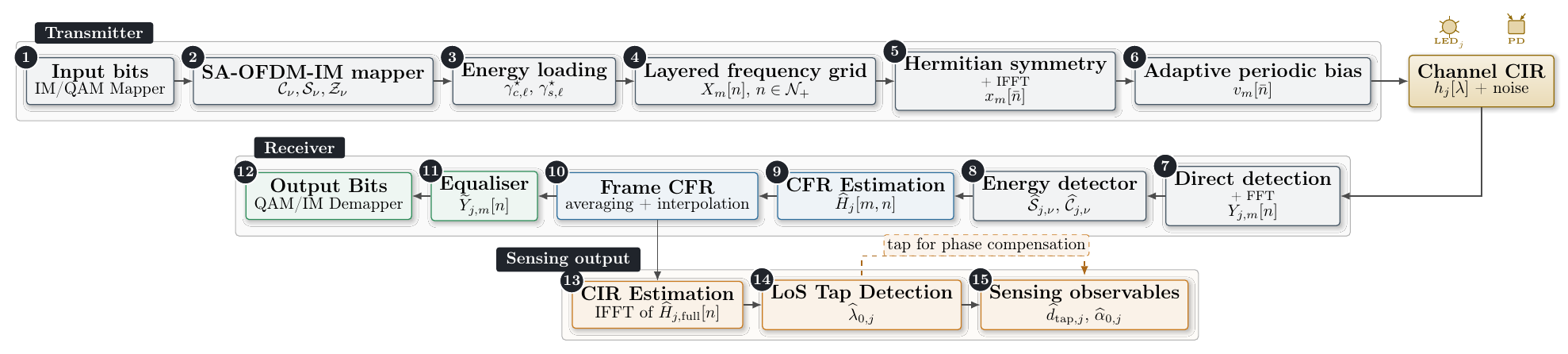}
    \caption{Proposed SA-OFDM-IM transceiver architecture for optical \ac{ICL}, including transmitter waveform generation, receiver-side subcarrier detection and \ac{CE}, and extraction of LoS delay and gain observables for sensing.}
\label{fig:sa_ofdm_im_transceiver}
\end{figure*}
\subsection{Transmitter Design}
Fig.~\ref{fig:sa_ofdm_im_transceiver} depicts the proposed \ac{SA}-\ac{OFDM}-\ac{IM} transceiver architecture. The waveform employs \(N\) subcarriers and \(N_{\mathrm{sym}}\) \ac{OFDM} symbols per frame indexed by \(m\in\{0,1,\ldots,N_{\mathrm{sym}}-1\}\), over a subcarrier spacing \(\Delta f\) that fixes the signal bandwidth \(B=N\Delta f\), \ac{OFDM} symbol duration \(T_{\mathrm{s}}=\frac{1}{\Delta f}\), and sampling interval \(T_{\mathrm{sam}}=\frac{T_{\mathrm{s}}}{N}\). Since \ac{IM/DD} transmission mandates a real-valued \ac{TD} waveform, Hermitian symmetry is enforced in the \ac{FD}, which restricts data-bearing modulation to the \ac{PF} subcarrier set \(\mathcal{N}_{+}=\{1,2,\ldots,\frac{N}{2}-1\}\), while the DC and Nyquist bins are nulled.

The proposed frame structure generalises the multilayer principle of \cite{wang2023clipping} to an \ac{IM}-enabled waveform by superimposing \(L\) \ac{FD} layers, each exploiting group-wise \ac{IM} to jointly embed communication symbols and sensing pilots. Layer \(\ell\) occupies the dyadically decimated subcarrier set \(\mathcal{U}_{\ell}=\{2^{\ell-1}+a2^{\ell}:a=0,1,\ldots,U_{\ell}-1\}\), \(\ell=1,\ldots,L\), with \(U_{\ell}=\frac{N}{2^{\ell+1}}\). The first layer thus coincides with the odd \ac{PF} subcarriers, as in \ac{ACO}-\ac{OFDM}, while successive layers occupy progressively sparser, higher-order decimations. Within each layer, \(U_{\ell}\) admissible subcarriers are partitioned into \(G_{\ell}\) disjoint groups of size \(g_{\ell}\), adaptively selected as \(g_{\ell}=\max\{g\in\{16,8,4\}:g\mid U_{\ell}\}\), so that \(G_{\ell}=\frac{U_{\ell}}{g_{\ell}}\). A layer is retained only if \(U_{\ell}\geq4\), the minimum group size admitting a non-trivial activation pattern. Moreover, since \(U_{\ell}=\frac{N}{2^{\ell+1}}\), this yields the usable-layer bound \(\ell\leq\log_2(\frac{N}{4})-1\), and hence \(L=\lfloor\log_2(\frac{N}{4})\rfloor-1\), with layers \(\ell>L\) excluded from \ac{IM} loading.

\begin{remark}
The adaptive group-size rule assigns coarser groups (\(g_\ell=16\)) to subcarrier-rich lower layers and finer groups (\(g_\ell=4\)) to subcarrier-sparse higher layers, maximising the combinatorial \ac{IM} payload \(\lfloor\log_2\binom{g_\ell}{k_\ell}\rfloor\), which grows with \(g_\ell\), wherever subcarrier availability permits. 
\end{remark}

Let \(\nu\triangleq(m,\ell,i)\) denote the symbol-layer-group tuple, \(i\in\{0,1,\ldots,G_{\ell}-1\}\), and \(\mathcal{R}_{\ell}\triangleq\{0,1,\ldots,g_{\ell}-1\}\) the intra-group tone index set. The \ac{PF} subcarrier mapping \(n_{\ell,i,r}=2^{\ell-1}+(ig_{\ell}+r)2^{\ell}\) resolves group position \(ig_{\ell}+r\) into an absolute bin index. The factor \(2^{\ell}\) enforces the layer-specific inter-subcarrier spacing, while the offset \(2^{\ell-1}\) guarantees \(n_{\ell,i,r}\equiv2^{\ell-1}\pmod{2^{\ell}}\). For each \(\nu\), \(\mathcal{R}_{\ell}\) is partitioned into disjoint communication, sensing-pilot, and null index subsets, \(\mathcal{C}_{\nu}\), \(\mathcal{S}_{\nu}\), and \(\mathcal{Z}_{\nu}\)\ of cardinalities \(k_{\ell}\), \(p_{\ell}\), and \(z_{\ell}=g_{\ell}-k_{\ell}-p_{\ell}\), respectively, such that \(\mathcal{R}_{\ell}=\mathcal{C}_{\nu}\,\dot\cup\,\mathcal{S}_{\nu}\,\dot\cup\,\mathcal{Z}_{\nu}\). \(\mathcal{C}_{\nu}\) is determined by the incoming \ac{IM} bits, such that, a group of \(g_{\ell}\) positions admits \(\binom{g_{\ell}}{k_{\ell}}\) candidate activation patterns, of which only \(2^{\lfloor\log_2\binom{g_{\ell}}{k_{\ell}}\rfloor}\) are retained to enable fixed-length binary indexing, deterministically enumerated and mapped to \(\mathcal{C}_{\nu}\) via Pascal-triangle-based combinatorial mapping \cite{azim2020enhanced}. \(\mathcal{S}_{\nu}\) is then drawn uniformly from the residual set \(\mathcal{R}_{\ell}\setminus\mathcal{C}_{\nu}\), with the remainder \(\mathcal{Z}_{\nu}=\mathcal{R}_{\ell}\setminus(\mathcal{C}_{\nu}\cup\mathcal{S}_{\nu})\) nulled.
\vspace{-2mm}
\begin{remark}
Fixing the index length to $\lfloor\log_2\binom{g_\ell}{k_\ell}\rfloor$ bits discards $\binom{g_\ell}{k_\ell}-2^{\lfloor\log_2\binom{g_\ell}{k_\ell}\rfloor}$ patterns. Which patterns are dropped is a design choice affecting the retained codebook's minimum pairwise Hamming distance. The lexicographic Pascal enumeration adopted is not claimed to be distance-optimal.
\end{remark}

Let \(\mathcal{A}\subset\mathbb{C}\) denote a unit-energy \(M_{\mathrm{QAM}}\)-ary \ac{QAM} alphabet, and let \(a_{\nu}(r)\in\mathcal{A}\) denote the constellation symbol assigned to tone \(r\), with \(\mathbb{E}\{|a_{\nu}(r)|^2\}=1\). Similarly, let \(s_{\nu}(r)\) denote a known unit-energy Golay pilot sequence, \(|s_{\nu}(r)|^2=1\). The \ac{FD} symbol assigned to tone \(r\in\mathcal{R}_{\ell}\) is given as:
\begin{equation}
d_{\nu}(r)
=
\begin{cases}
\sqrt{\gamma^\star_{\mathrm c,\ell}}\,a_{\nu}(r), & r\in\mathcal{C}_{\nu},\\
\sqrt{\gamma^\star_{\mathrm s,\ell}}\,s_{\nu}(r), & r\in\mathcal{S}_{\nu},\\
0, & r\in\mathcal{Z}_{\nu},
\end{cases}
\label{eq:local_isac_symbol}
\end{equation}
where \(\gamma^\star_{\mathrm c,\ell}\) and \(\gamma^\star_{\mathrm s,\ell}\) are the layer-\(\ell\) optimal per-subcarrier energy-allocation coefficients analysed in Section~\ref{sec2}-C, with \(\gamma^\star_{\mathrm s,\ell}>\gamma^\star_{\mathrm c,\ell}\) imposed by design.

Mapping \(d_{\nu}(r)\) onto its assigned bin, \(X_m[n_{\ell,i,r}]=d_{\nu}(r)\), the composite \ac{FD} symbol over all layers and groups is:
\begin{equation}
X_m[n]
=
\sum_{\ell=1}^{L}
\sum_{i=0}^{G_{\ell}-1}
\sum_{r\in\mathcal{R}_{\ell}}
d_{\nu}(r)\delta_\mathrm{K}[n-n_{\ell,i,r}],
\label{eq:positive_frequency_grid}
\end{equation}
for \(n\in\mathcal{N}_{+}\). Hermitian symmetry is then imposed, \(X_m[N-n]=X_m^{\ast}[n]\) for \(1\leq n\leq\frac{N}{2}-1\), with \(X_m[0]=X_m[\frac{N}{2}]=0\), extending \(X_m[n]\) over the full grid \(n=0,1,\ldots,N-1\). The corresponding real-valued, bipolar \ac{TD} signal follows from the inverse \ac{DFT} as:
\begin{equation}
x_m[\bar{n}]
=
\frac{1}{\sqrt{N}}
\sum_{n=0}^{N-1}
X_m[n]e^{\frac{\jmath2\pi n\bar{n}}{N}},
\label{eq:time_domain_signal}
\end{equation}
for \(\bar{n}=0,1,\ldots,N-1\), where \(\jmath = \sqrt{-1}\). To enforce the non-negativity constraint of \ac{IM/DD}, an adaptive periodic bias is added following \cite{wang2023clipping}. The symbol is partitioned into \(2^L\) sub-blocks of length \(N_{\mathrm{seg}}=\frac{N}{2^L}\), \(x_{m,\tau}[\beta]=x_m[\beta+\tau N_{\mathrm{seg}}]\), \(\beta=0,\ldots,N_{\mathrm{seg}}-1\), \(\tau=0,\ldots,2^L-1\), and the minimum bias satisfying non-negativity at each intra-segment position is computed as \(b_{m,0}[\beta]=-\min_{0\leq \tau\leq2^L-1}x_{m,\tau}[\beta]\), periodically extended as \(b_m[\bar{n}]=b_{m,0}[\bar{n}\bmod N_{\mathrm{seg}}]\). The transmitted \ac{OFDM} symbol is therefore given as:
\begin{equation}
v_m[\bar{n}]
=
x_m[\bar{n}]+b_m[\bar{n}],
\label{eq:unipolar_waveform}
\end{equation}
satisfying \(v_m[\bar{n}]\geq0~\forall~\bar{n}\). \(v_m[\bar{n}]\) is transmitted following the insertion of a \ac{CP} of length \(N_{\mathrm{CP}}\).

\begin{remark}
The \(N_{\mathrm{seg}}\)-periodicity of \(b_m[\bar{n}]\) confines its spectral support to bins that are integer multiples of \(2^L\) \cite[Eqs.~(8)-(9)]{wang2023clipping}. Since every occupied subcarrier \(n_{\ell,i,r}\) satisfies \(n_{\ell,i,r}\equiv2^{\ell-1}\pmod{2^{\ell}}\) for \(\ell\leq L\) and is therefore never a multiple of \(2^L\), the bias tone falls on unmodulated bins, leaving both the communication symbols and the sensing pilots undistorted.
\end{remark}
\vspace{-2mm}
\subsection{Receiver Processing and Sensing-Observable Extraction}
Following direct detection and \ac{CP} removal, the received discrete-time signal associated with \ac{LED} \(j\) is the linear convolution of the transmitted waveform \(v_m[\bar{n}]\) with the discrete-time \ac{CIR} \(h_j[\lambda]\), corrupted by additive noise:
\begin{equation}
y_{j,m}[\bar{n}]= \sum_{\lambda=0}^{L_h-1}
h_j[\lambda]\,v_m[\bar{n}-\lambda]
+ w_{j,m}[\bar{n}],
\label{eq:received_time_domain_signal}
\end{equation}
where \(w_{j,m}[\bar{n}]\) is zero-mean \ac{AWGN} sample. Applying an \(N\)-point \ac{DFT} gives:
\begin{equation}
Y_{j,m}[n] = H_j[n]V_m[n] + W_{j,m}[n],
\label{eq:received_frequency_domain_signal}
\end{equation}
for \(n=0,1,\ldots,N-1\), where \(H_j[n]\) is the \ac{CFR} between \ac{LED} \(j\) and the \ac{PD}, \(V_m[n]\) is the transmitted \ac{FD} symbol, and \(W_{j,m}[n]\) is the \ac{FD} noise term.

Since the true subcarrier partition \(\{\mathcal{C}_\nu,\mathcal{S}_\nu,\mathcal{Z}_\nu\}\) is not known \textit{a priori} at the receiver, and no channel estimate is yet available to enable coherent detection, sensing subcarrier recovery must precede equalisation which is performed directly on the received energy. For each tuple \(\nu\), the receiver evaluates the per-tone energy metric \(
\Gamma_{j,\nu}(r) = \left|Y_{j,m}[n_{\ell,i,r}] \right|^2\), for \(r\in\mathcal{R}_{\ell}\). Because the sensing pilots are allocated strictly higher per-subcarrier energy than the communication symbols by construction, the imposed power-domain separation renders the sensing and communication subcarriers distinguishable by rank order alone. Let \(\pi_{j,\nu}(e)\) denote the local tone index associated with the \(e\)-th largest value of \(\Gamma_{j,\nu}(r)\) within tuple \(\nu\). The \(p_{\ell}\) highest-energy tones are declared as the sensing-pilots,
\(\widehat{\mathcal{S}}_{j,\nu}
=\left\{\pi_{j,\nu}(1),\ldots,\pi_{j,\nu}(p_{\ell})\right\}\), and the next \(k_{\ell}\) highest-energy tones are declared as the communication subcarriers,
\(\widehat{\mathcal{C}}_{j,\nu}=\left\{\pi_{j,\nu}(p_{\ell}+1),\ldots,\pi_{j,\nu}(p_{\ell}+k_{\ell})\right\}\). This energy-based subcarrier detection lowers receiver complexity from the combinatorial joint search $\mathcal{O}\!\left(2^{\lfloor\log_2\binom{g_\ell}{k_\ell}\rfloor}M_{\mathrm{QAM}}^{k_\ell}\right)$ required by exhaustive \ac{ML} detection to a single sort of $g_\ell$ energy values, i.e., $\mathcal{O}(g_\ell\log g_\ell)$. This reduction is enabled by the energy-allocation constraint $\gamma^\star_{\mathrm s,\ell}>\gamma^\star_{\mathrm c,\ell}$ so that, with high probability, the energy of every sensing tone outranks every communication tone in $\Gamma_{j,\nu}(r)$.

For each detected sensing-pilot subcarrier, the known pilot symbol enables direct \ac{LS} \ac{CFR} estimation. Using the sensing energy coefficient \(\gamma_{\mathrm{s},\ell}^{\star}\), the estimate for \(r\in\widehat{\mathcal{S}}_{j,\nu}\) is:
\begin{equation}
\widehat{H}_{j,m}[n_{\ell,i,r}]=
\frac{Y_{j,m}[n_{\ell,i,r}]}{
\sqrt{\gamma_{\mathrm{s},\ell}^{\star}}\,s_{\nu}(r)}.
\label{eq:ls_channel_estimate}
\end{equation}

Because the \ac{IM} activation pattern varies from one tuple \(\nu\) to the next, the sensing-pilot locations are not fixed across the frame.
The receiver therefore exploits this temporal diversity by aggregating channel observations over the entire frame of \(N_{\mathrm{sym}}\) symbols. The frame-level sensed \ac{PF} set is given as:
\begin{equation}
\mathcal{P}_{j,\mathrm{fr}}=\bigcup_{m=0}^{N_{\mathrm{sym}}-1}\bigcup_{\ell=1}^{L}\bigcup_{i=0}^{G_{\ell}-1}
\left\{n_{\ell,i,r}:r\in\widehat{\mathcal{S}}_{j,\nu}
\right\}.
\label{eq:frame_sensing_support}
\end{equation}
For each sensed subcarrier \(n\), the associated observation set is \(\mathcal{M}_{j,n}
=\left\{\nu= (m,\ell,i),r:n_{\ell,i,r}=n,\;r\in\widehat{\mathcal{S}}_{j,\nu}\right\}\), and repeated estimates at the same bin are combined via sample averaging as:
\begin{equation}
\overline{H}_j[n]= \frac{1}{|\mathcal{M}_{j,n}|}\!\!
\sum_{(\nu,r)\in\mathcal{M}_{j,n}}\!\!\!\!\!\!\!\!
\widehat{H}_{j,m}[n_{\ell,i,r}],
\qquad
|\mathcal{M}_{j,n}|\geq1 .
\label{eq:averaged_cfr}
\end{equation}

\begin{remark}
Because the noise in \eqref{eq:ls_channel_estimate} is uncorrelated across \ac{OFDM} symbols, averaging the \(|\mathcal{M}_{j,n}|\) independent estimates of a subcarrier suppresses the estimator noise variance by a factor of \(|\mathcal{M}_{j,n}|\). This gain arises directly from the layered \ac{IM} structure, requiring no training overhead.
\end{remark}

The frame-level \ac{PF} \ac{CFR} estimate follows as:
\begin{equation}
\widehat{H}_{j,+}[n]
=
\begin{cases}
\overline{H}_j[n],
& n\in\mathcal{P}_{j,\mathrm{fr}},
\\[2pt]
\text{interpolated value},
& n\notin\mathcal{P}_{j,\mathrm{fr}},
\end{cases},
\label{eq:positive_frequency_cfr_estimate}
\end{equation}
for \(n = 1,2,\cdots,\frac{N}{2}-1\), where linear interpolation fills subcarriers that remain unobserved over a finite frame length. Imposing Hermitian symmetry on \(\widehat{H}_{j,+}[n]\) yields the full-band estimate \(\widehat{H}_{j,\mathrm{full}}[n]\), from which the corresponding \ac{CIR} estimate follows via inverse \ac{DFT} as:
\begin{equation}
\widehat{h}_{j,\mathrm{fr}}[\lambda]
=
\frac{1}{\sqrt{N}}
\sum_{n=0}^{N-1}
\widehat{H}_{j,\mathrm{full}}[n]
e^{\frac{\jmath 2\pi n\lambda}{N}},
\label{eq:estimated_cir}
\end{equation}
for \(\lambda=0,1,\ldots,N-1\), where \(\lambda\) is the delay-tap index.

Moreover, the received \ac{PF} subcarriers are equalised via zero-forcing division by the frame-level \ac{CFR} estimate as:
\begin{equation}
\widetilde{Y}_{j,m}[n]
=
\frac{Y_{j,m}[n]}{\widehat{H}_{j,+}[n]},
\label{eq:equalised_received_symbol}
\end{equation}
for \(n = 1,2,\cdots,\frac{N}{2}-1\). For each tuple \(\nu\), the equalised observations on the detected communication subcarriers are extracted as \(\widetilde{d}_{j,\nu}(r)=
\widetilde{Y}_{j,m}[n_{\ell,i,r}]\), for \(r\in\widehat{\mathcal{C}}_{j,\nu}\). Since the transmitted symbol on an active communication subcarrier is \(\sqrt{\gamma_{\mathrm c,\ell}^{\star}}a_{\nu}(r)\), the corresponding de-normalised \ac{QAM} observation is
\(\widehat{a}_{j,\nu}(r)=
\frac{\widetilde{d}_{j,\nu}(r)}{\sqrt{\gamma_{\mathrm c,\ell}^{\star}}}\), and the recovered \ac{QAM} symbol follows from minimum-Euclidean-distance detection over the constellation alphabet,
\(\widehat{a}_{j,\nu}^{\mathrm{det}}(r)
=\arg\min_{a\in\mathcal{A}}
\left|\widehat{a}_{j,\nu}(r)-a
\right|^2\), for \(r\in\widehat{\mathcal{C}}_{j,\nu}\). The detected subcarriers \(\widehat{\mathcal{C}}_{j,\nu}\) simultaneously serves as the estimated \ac{IM} activation pattern for tuple \(\nu\) and is mapped back to its bit representation via inverse Pascal-triangle-based combinatorial demapping \cite{azim2020enhanced}. The recovered information bits for each group are thus formed by concatenating the demapped index bits with the bits recovered from the detected \ac{QAM} symbols.

The receiver only registers integer-spaced delays \cite{xu2025integrated}. Since the \ac{LoS} component constitutes the shortest propagation path between \ac{LED} \(j\) and the \ac{PD}, it manifests as the first significant arrival in the estimated \ac{CIR}. The estimated \ac{LoS} tap is therefore obtained as:
\begin{equation}
\widehat{\lambda}_{0,j}
=
\min
\left\{
\lambda\in\mathcal{L}_{\mathrm{search}}:
\left|\widehat{h}_{j,\mathrm{fr}}[\lambda]\right|
\geq
\varkappa_j
\right\},
\label{eq:first_arrival_tap_detector}
\end{equation}
where \(\varkappa_j\) is the detection threshold, \(|\cdot|\) is the absolute operator, and the search is restricted to the delay window \(\mathcal{L}_{\mathrm{search}}=\{0,1,\ldots,\lambda_{\max}\}\), with \(\lambda_{\max}\) determined by the maximum admissible \ac{LED}-\ac{PD} distance under the room geometry and clipped to \(N-1\). The threshold is set adaptively relative to the estimated \ac{CIR} magnitude as \(\varkappa_j =
0.2 \max_{\lambda\in\mathcal{L}_{\mathrm{search}}}
\left|\widehat{h}_{j,\mathrm{fr}}[\lambda]\right|\). The threshold \(\varkappa_j\) trades off two failure modes, i.e., an excessively high threshold risks missing a weak but genuine \ac{LoS} tap in deeply attenuated links, biasing \(\widehat{\lambda}_{0,j}\) toward a later, spurious arrival, whereas an excessively low threshold risks falsely triggering on residual noise, biasing \(\widehat{\lambda}_{0,j}\) toward an earlier tap. Normalising \(\varkappa_j\) relative to the per-link \ac{CIR} peak, rather than fixing an absolute threshold, makes the detector less sensitive to the absolute \ac{LED}-\ac{PD} link gains encountered over the room geometry.

The detected tap \(\widehat{\lambda}_{0,j}\) in \eqref{eq:first_arrival_tap_detector} is converted into a delay-based range estimate as:
\vspace{-2.4mm}
\begin{equation}
\widehat{d}_{0,j}
=
\frac{c_0\widehat{\lambda}_{0,j}}{B},
\label{eq:tap_range_from_detected_lambda}
\end{equation}
which is used for delay-based localisation. In addition, \(\widehat{\lambda}_{0,j}\) enables extraction of the \ac{LoS} gain from the estimated \ac{CFR}. Under \ac{LoS}-dominant propagation, the \ac{CFR} for \ac{LED} \(j\) reduces to a single complex exponential as:
\begin{equation}
H_{j,+}[n]
\approx
\alpha_{0,j}
e^{-\frac{\jmath2\pi n\lambda_{0,j}}{N}},
\label{eq:single_tap_cfr_model}
\end{equation}
for \(n=1,2,\cdots,\frac{N}{2}-1\). Multiplying the estimated \ac{CFR} by the conjugate phase ramp \(e^{\frac{\jmath2\pi n\widehat{\lambda}_{0,j}}{N}}\) compensates the delay-induced linear phase rotation across frequency, coherently aligning the \ac{LoS} contribution. The corresponding gain estimate follows from averaging the phase-compensated \ac{CFR} samples over \(\mathcal{N}_+\) as:
\begin{equation}
\widehat{\alpha}_{0,j}
=\left|
\frac{1}{|\mathcal{N}_+|}
\sum_{n\in\mathcal{N}_+}
\widehat{H}_{j,+}[n]
e^{\frac{\jmath2\pi n\widehat{\lambda}_{0,j}}{N}}
\right|.
\label{eq:alpha_hat}
\end{equation}
%

After phase compensation, the \ac{LoS} term has constant, zero phase across \(n\), so it adds up constructively. This gain estimator is unbiased under the ideal single-tap \ac{LoS} model \eqref{eq:single_tap_cfr_model} in the absence of estimation errors. 
%
\vspace{-4mm}
\subsection{Energy Allocation Optimisation}
As established in Section~\ref{sec2}-B, the recovery of \ac{IM} activation pattern and the sensing-pilots is performed by an energy detector rather than by exhaustive \ac{ML} detection. The viability of this low-complexity detector depends on the fact that the sensing and communication subcarriers must remain separable in the power domain. This separability is enforced through a sensing-to-communication energy margin \(R_{\min}\), such that, for every layer \(\ell\), we have \(
\frac{\gamma_{\mathrm s,\ell}}{\gamma_{\mathrm c,\ell}}\geq R_{\min}=10^{\Delta_{\mathrm s}/10}\), where \(R_{\min}\geq1\) and \(\Delta_{\mathrm s}\) denotes the minimum sensing margin expressed in dB.


To parameterise this trade-off, the reference per-subcarrier energies are initialised as \(\{\gamma_{\mathrm c}^{\mathrm{ref}},\gamma_{\mathrm s}^{\mathrm{ref}}\}=\{1,1\}\), yielding the reference active-group energy of layer \(\ell\) as \(E_{\ell}^{\mathrm{ref}}=k_{\ell}\gamma_{\mathrm c}^{\mathrm{ref}}+p_{\ell}\gamma_{\mathrm s}^{\mathrm{ref}}\). Let \(\mu\in(0,1)\) denote the fraction of \(E_{\ell}^{\mathrm{ref}}\) apportioned to sensing. The resulting per-subcarrier energies are \(
\gamma_{\mathrm s,\ell}(\mu)=\frac{\mu E_{\ell}^{\mathrm{ref}}}{p_{\ell}}\) and 
\(\gamma_{\mathrm c,\ell}(\mu)=\frac{(1-\mu)E_{\ell}^{\mathrm{ref}}}{k_{\ell}}\) which satisfy \(E_{\ell}^{\mathrm{ref}} =k_{\ell}\gamma_{\mathrm c,\ell}(\mu)+p_{\ell}\gamma_{\mathrm s,\ell}(\mu)\) for all \(\mu\). The parameterisation is thus \textit{energy-conserving}, so that varying \(\mu\) redistributes the transmit energy, and all operating points remain directly comparable at fixed average power. Although a single scalar \(\mu\) is applied across all layers, the induced energies, \(\gamma_{\mathrm c,\ell}(\mu)\) and \(\gamma_{\mathrm s,\ell}(\mu)\) remain layer-dependent, since \(k_{\ell}\), \(p_{\ell}\), and \(E_{\ell}^{\mathrm{ref}}\) are determined by the adaptive group configuration presented in Section~\ref{sec2}-A. After straightforward manipulation, we obtain \(\frac{\gamma_{\mathrm s,\ell}(\mu)}{\gamma_{\mathrm c,\ell}(\mu)}=\frac{\mu k_{\ell}}{(1-\mu)p_{\ell}}\), so that feasibility requires \(
\mu\geq\max_{1\leq\ell\leq L}\frac{R_{\min}p_{\ell}}{k_{\ell}+R_{\min}p_{\ell}}\).


The optimal allocation fraction is then obtained as:
\begin{equation}
\mu^{\star}=\arg\min_{\mu\in\mathcal{M}}J_{\mathrm E}(\mu),
\label{eq:energy_allocation_problem}
\end{equation}
where the feasible set is \(\mathcal{M}=\{\mu\in(0,1):\frac{\gamma_{\mathrm s,\ell}(\mu)}{\gamma_{\mathrm c,\ell}(\mu)}\geq R_{\min}~\forall\ell\}\), and the bi-objective cost is defined as:
\begin{equation}
J_{\mathrm E}(\mu)
=
w_{\mathrm c}\log_{10}
\left(
1+
\frac{F_{\mathrm c}(\mu)}
{F_{\mathrm c}^{\mathrm{tar}}}
\right)
+
w_{\mathrm s}\log_{10}
\left(
1+
\frac{F_{\mathrm s}(\mu)}
{F_{\mathrm s}^{\mathrm{tar}}}
\right).
\label{eq:normalised_log_objective}
\end{equation}
To prevent the operating point from being over-fitted to a single \ac{SNR}, both performance functionals are averaged over a representative \ac{SNR} set \(\mathcal{S}_{\mathrm{opt}}\), namely \(F_{\mathrm c}(\mu)=\frac{1}{|\mathcal{S}_{\mathrm{opt}}|}\sum_{\xi\in\mathcal{S}_{\mathrm{opt}}}\operatorname{BER}(\mu;\xi)\) and \(F_{\mathrm s}(\mu)=\frac{1}{|\mathcal{S}_{\mathrm{opt}}|}\sum_{\xi\in\mathcal{S}_{\mathrm{opt}}}\operatorname{RMSE}_{\mathrm{pos}}(\mu;\xi)\), where \(\xi\) indexes an \ac{SNR} value in \(\mathcal{S}_{\mathrm{opt}}\). The weights \(w_{\mathrm c}\) and \(w_{\mathrm s}\) set the relative priority of the communication and sensing criteria, while the target levels \(F_{\mathrm c}^{\mathrm{tar}}\) and \(F_{\mathrm s}^{\mathrm{tar}}\) render the two dimensionally incommensurate metrics comparable prior to weighting, since \(F_{\mathrm c}(\mu)\) is a dimensionless error probability whereas \(F_{\mathrm s}(\mu)\) is a positioning error expressed in metres.

\begin{remark}
The logarithmic compression in \eqref{eq:normalised_log_objective} is essential to the balance of the formulation. The \ac{BER} varies over several orders of magnitude across \(\mathcal{M}\), whereas the localisation \ac{RMSE} varies far more gradually. Since the delay-based range measurements possess a quantisation floor of \(\frac{\Delta d}{2}\), which ultimately limits delay-only localisation accuracy. Under a linear combination, the communication term would consequently dominate the objective irrespective of \(w_{\mathrm c}\) and \(w_{\mathrm s}\), reducing the bi-objective design to a de facto \ac{BER} minimisation. Compressing both terms equalises their dynamic ranges, so that the weights recover their intended interpretation.
\end{remark}


Given \(\mu^{\star}\), the optimised per-subcarrier energies are \(\gamma_{\mathrm s,\ell}^{\star}=\frac{\mu^{\star}E_{\ell}^{\mathrm{ref}}}{p_{\ell}}\) and \(\gamma_{\mathrm c,\ell}^{\star}=\frac{(1-\mu^{\star})E_{\ell}^{\mathrm{ref}}}{k_{\ell}}\), with the realised sensing-to-communication ratio \(R_{\ell}^{\star}=\frac{\mu^{\star}k_{\ell}}{(1-\mu^{\star})p_{\ell}}\geq R_{\min}\). For a given deployment scenario, \eqref{eq:energy_allocation_problem} is solved once using \(g_{\ell}\), \(k_{\ell}\), \(p_{\ell}\), \(F_{\mathrm c}^{\mathrm{tar}}\), \(F_{\mathrm s}^{\mathrm{tar}}\), \(w_{\mathrm c}\), \(w_{\mathrm s}\), and \(\mathcal{S}_{\mathrm{opt}}\).  The resulting \(\mu^{\star}\) is thereafter treated as a fixed, scenario-specific design parameter rather than as an adaptive quantity updated during transmission.

\begin{remark}
\(\mu^{\star}\) is optimal only in the average sense over \(\mathcal{S}_{\mathrm{opt}}\) and the assumed room geometry, not re-optimised for instantaneous link conditions. Moreover, regarding \(\mu^{\star}\), we observe two important properties. First, \(J_{\mathrm E}(\mu)\) is comparatively flat near its minimiser, so moderate mismatch in \(\mu\) costs little. Second, the sensing term \(F_{\mathrm s}(\mu)\) grows less sensitive to \(\mu\) as bandwidth \(B\) increases, since the delay-based \ac{RMSE} approaches its quantisation floor and stops rewarding extra sensing energy, leaving the optimisation effectively communication-driven, which is why \(\mu^{\star}\) stays sensing-dominant but never sensing-exclusive.
\end{remark}
\vspace{-5mm}
\section{Delay-Aided RSS Localisation}
\label{sec:delay_rss_ranging_position}
In this section, we develop a delay-aided \ac{RSS}-based localisation framework that exploits two complementary observables recovered from the estimated \ac{CIR}, that are, the detected \ac{LoS} delay tap \(\widehat{\lambda}_{0,j}\) and the phase-compensated \ac{LoS} gain \(\widehat{\alpha}_{0,j}\). The delay observable, extracted purely from the tap index, serves as a direct time-of-flight proxy for the \ac{LED}-to-\ac{PD} propagation distance and is inherently insensitive to the \ac{PD} surface normal \(\vect{n}_{\mathrm D}\) and to optical-gain calibration. The gain observable, by contrast, is governed jointly by the inverse-square distance law, the Lambertian emergence projection, and the \ac{PD} incidence projection, so that a single scalar measurement couples position and orientation. The delay therefore yields a range-only estimate, whereas the gain additionally encodes the emission and incidence angles, so the two observables carry largely non-redundant and complementary geometric information. Delay-based ranging and multilateration consequently depend only on \(\{\widehat{\lambda}_{0,j},\vect{c}_{\mathrm S,j}\}\) and are independent of \(\widehat{\alpha}_{0,j}\), whereas the proposed \ac{RSS} stage treats the delay as a coarse anchor and refines it by enforcing amplitude consistency. The resulting estimator may thus be interpreted as a delay-constrained amplitude-fitting problem, in which the delay branch supplies coarse distance and the gain branch supplies sub-quantisation refinement together with orientation observability.

For a sampling interval \(T_{\mathrm{sam}}\), the \ac{LoS} propagation delay associated with the detected tap \(\widehat{\lambda}_{0,j}\) is:
\begin{equation}
\widehat{\tau}_{0,j}=\widehat{\lambda}_{0,j}T_{\mathrm{sam}}=\frac{\widehat{\lambda}_{0,j}}{B}
\;\approx\;
\tau_{0,j}=\frac{d_{0,j}}{c_0},
\label{eq:estimated_los_delay_from_tap}
\end{equation}
where the approximation reflects the fact that \(\widehat{\tau}_{0,j}\) is constrained to the discrete sampling lattice and therefore cannot, in general, coincide exactly with the continuous-valued true delay \(\tau_{0,j}\). The associated range estimate \(\widehat{d}_{0,j}\) in \eqref{eq:tap_range_from_detected_lambda} is by construction unaffected by \(\vect{n}_{\mathrm D}\), by the incidence angle, or by \(\widehat{\alpha}_{0,j}\). Its accuracy is instead limited exclusively by delay quantisation. Since the channel is observed on a uniform sampling grid, two propagation paths whose delays fall within the same bin are unresolvable from the tap index alone. The physical range interval subtended by one delay bin is \(d=c_0T_{\mathrm{sam}}=\frac{c_0}{B}\), so that the delay-derived range exhibits a deterministic quantisation error bounded by \(\frac{\Delta d}{2}\) in magnitude.

\begin{remark}
Higher \ac{SNR} only improves tap-detection reliability in \eqref{eq:first_arrival_tap_detector} and cannot reduce \(\Delta d\). Thus,  \(\Delta d\) improves only linearly with \(B\), leaving a structural, irreducible error floor of \(\frac{\Delta d}{2}\) even in a noiseless scenario. \(B\) is bounded by sampling and subcarrier-spacing choices and by the \ac{LED}/\ac{PD} opto-electronic bandwidths. Conventional off-the-shelf \acp{LED} typically offer modulation bandwidths of only a few \si{\mega\hertz} \cite{xu2025microled}, whereas emerging micro-\ac{LED} devices have been demonstrated with bandwidths extending into the hundreds of \si{\mega\hertz} to a few \si{\giga\hertz} \cite{liao2026photodetector}. Consequently, the range resolution becomes a relevant parameter impacting the localisation performance.
\end{remark}

Let \(\mathcal{J}_{\mathrm v}\subseteq\{1,2,\ldots,J\}\) denote the set of usable \ac{LED} links, comprising those links for which both a reliable \ac{LoS} tap and a valid gain observation are available. \(\mathcal{J}_{\mathrm v}\) excludes \acp{LED} whose \ac{LoS} paths are blocked, that fall outside the \ac{PD} \ac{FoV}, whose \ac{LoS} taps fail to exceed the detection threshold \(\varkappa_j\), or whose gain estimates are corrupted by unresolved multipath. The corresponding delay-based range estimates are \(\{\widehat{d}_{0,j}\}_{j\in\mathcal{J}_{\mathrm v}}\). For a candidate \ac{PD} position \(\vect{c}=[x,y,z]^{\mathrm T}\), the geometric distance to \ac{LED} \(j\) is \(d_j(\vect{c})=\norm{\vect{c}-\vect{c}_{\mathrm S,j}}\), and the delay-based position estimate follows from multilateration as:
\begin{equation}
\widehat{\vect{c}}_{\mathrm D}^{(\mathrm{del})}
=\operatorname*{arg\,min}_{\vect{c}\in\mathcal{B}}\sum_{j\in\mathcal{J}_{\mathrm v}}\left(d_j(\vect{c})-\widehat{d}_{0,j}\right)^2,
\label{eq:delay_multilateration}
\end{equation}
over the admissible search region \(\mathcal{B}=\{\vect{c}=[x,y,z]^{\mathrm T}:0\leq x\leq L_x,\;0\leq y\leq L_y,\;z_{\min}\leq z\leq z_{\max}\}\), where \(L_x\) and \(L_y\) are the room dimensions and \(\{z_{\min},z_{\max}\}\) bound the admissible \ac{PD} height within a room of height \(L_z\). 
The estimate \eqref{eq:delay_multilateration} is nonetheless coarse, since each residual is formed against a quantised range measurement.

For \ac{RSS}-based localisation, the same candidate position \(\vect{c}\) parameterises the link geometry. The unit direction vector from \ac{LED} \(j\) towards \(\vect{c}\) is \(\vect{u}_{\mathrm{SD},j}(\vect{c})=\frac{\vect{c}-\vect{c}_{\mathrm S,j}}{d_j(\vect{c})}\), with reverse direction \(\vect{u}_{\mathrm{DS},j}(\vect{c})=-\vect{u}_{\mathrm{SD},j}(\vect{c})\), where the former determines the \ac{LED} emergence angle and the latter the \ac{PD} incidence angle. The emergence-angle cosine is rectified as \(\varsigma_{t,j}(\vect{c})=\max(\vect{u}_{\mathrm{SD},j}^{\mathrm T}(\vect{c})\vect{n}_{\mathrm S,j},0)\), so that candidate positions lying outside the radiating hemisphere of \ac{LED} \(j\) contribute no \ac{LoS} power. Collecting every position-dependent factor of the \ac{LoS} gain model \eqref{eq:los_gain} except the \ac{PD} incidence projection yields the orientation-independent kernel:
\begin{equation}
\kappa_j(\vect{c})=\frac{(q+1)A_{\mathrm{PD}}g_{\mathrm{of}}g_{\mathrm{oc}}}{2\pi d_j^2(\vect{c})}\varsigma_{t,j}^{q}(\vect{c}).
\label{eq:rss_candidate_kappa}
\end{equation}
%

For a candidate pair \((\vect{c},\vect{n})\), the incidence projection on link \(j\) is \(\varsigma_{r,j}(\vect{c},\vect{n})=\vect{u}_{\mathrm{DS},j}^{\mathrm T}(\vect{c})\vect{n}\), and the predicted \ac{LoS} gain is:
\begin{equation}
\alpha_{j}^{(\mathrm{pred})}(\vect{c},\vect{n})
=
\kappa_j(\vect{c})
\vect{u}_{\mathrm{DS},j}^{\mathrm T}(\vect{c})\vect{n},
\label{eq:rss_predicted_gain}
\end{equation}
subject to the physical validity conditions \(\norm{\vect{n}}=1\) and \(\vect{u}_{\mathrm{DS},j}^{\mathrm T}(\vect{c})\vect{n}\geq\cos(\Psi_{\mathrm{FoV}})\), the latter guaranteeing that \ac{LED} \(j\) lies within the \ac{FoV} of the candidate receiver. Stacking \eqref{eq:rss_predicted_gain} over all usable links yields the compact linear form:
\begin{equation}
\boldsymbol{\alpha}^{(\mathrm{pred})}(\vect{c},\vect{n})
=
\vect{A}(\vect{c})\vect{n},
\label{eq:rss_predicted_gain_vector}
\end{equation}
where \(\vect{A}(\vect{c})\in\mathbb{R}^{|\mathcal{J}_{\mathrm v}|\times3}\) is defined row-wise by \([\vect{A}(\vect{c})]_{a,:}=\kappa_{j_a}(\vect{c})\vect{u}_{\mathrm{DS},j_a}^{\mathrm T}(\vect{c})\), for \(a=1,\ldots,|\mathcal{J}_{\mathrm v}|\) and \(j_a\in\mathcal{J}_{\mathrm v}\), and the measured gain vector is \(\widehat{\boldsymbol{\alpha}}=[\widehat{\alpha}_{0,j_1},\ldots,\widehat{\alpha}_{0,j_{|\mathcal{J}_{\mathrm v}|}}]^{\mathrm T}\).


To treat the known- and unknown-orientation cases, we define the admissible normal set \(\mathcal{N}(\vect{c})\). When the instantaneous \ac{PD} orientation is known, \(\mathcal{N}(\vect{c})=\{\vect{n}_{\mathrm D}\}\) provided \(\vect{n}_{\mathrm D}\) satisfies the \ac{FoV} constraint for every \(j\in\mathcal{J}_{\mathrm v}\), and \(\mathcal{N}(\vect{c})=\varnothing\) otherwise, with \(\varnothing\) denoting the empty set. When the orientation is unknown, \(\mathcal{N}(\vect{c})\) comprises all unit-norm vectors satisfying \(\vect{u}_{\mathrm{DS},j}^{\mathrm T}(\vect{c})\vect{n}\geq\cos(\Psi_{\mathrm{FoV}})\) for every \(j\in\mathcal{J}_{\mathrm v}\), i.e., the intersection of the unit sphere with \(|\mathcal{J}_{\mathrm v}|\) half-spaces. The candidate-conditioned normal is then obtained as:
\begin{equation}
\widehat{\vect{n}}(\vect{c})
=
\operatorname*{arg\,min}_{\vect{n}\in\mathcal{N}(\vect{c})}
\norm{
\vect{A}(\vect{c})\vect{n}
-
\widehat{\boldsymbol{\alpha}}
}^{2},
\label{eq:rss_candidate_normal_general}
\end{equation}
which reduces to \(\widehat{\vect{n}}(\vect{c})=\vect{n}_{\mathrm D}\) in the known-orientation case and returns the \ac{FoV}-feasible unit normal of best amplitude fit otherwise. Candidate positions with \(\mathcal{N}(\vect{c})=\varnothing\) are declared infeasible. The associated predicted gain vector is:
\begin{equation}
\widehat{\boldsymbol{\alpha}}^{(\mathrm{pred})}(\vect{c})
=
\vect{A}(\vect{c})
\widehat{\vect{n}}(\vect{c}),
\label{eq:rss_candidate_predicted_gain_vector}
\end{equation}
so that an identical prediction model governs both orientation cases, differing only in the cardinality of \(\mathcal{N}(\vect{c})\).

Consistency between the predicted and measured gain patterns is quantified by the log-domain residual as:
\begin{equation}
J_{\mathrm{RSS}}(\vect{c})
=
\sum_{a=1}^{|\mathcal{J}_{\mathrm v}|}
\left[
\log\!\left(
\widehat{\alpha}_{a}^{(\mathrm{pred})}(\vect{c})
+
\epsilon_{\alpha}
\right)
-
\log\!\left(
\widehat{\alpha}_{0,j_a}
+
\epsilon_{\alpha}
\right)
\right]^2,
\label{eq:rss_position_log_cost}
\end{equation}
where \(\widehat{\alpha}_{a}^{(\mathrm{pred})}(\vect{c})\) is the \(a\)-th entry of \(\widehat{\boldsymbol{\alpha}}^{(\mathrm{pred})}(\vect{c})\), \(\epsilon_{\alpha}>0\) is a numerical regularisation constant guarding against vanishing gains, and infeasible candidates are assigned \(J_{\mathrm{RSS}}(\vect{c})=+\infty\).


The \ac{RSS} position estimate follows as:
\begin{equation}
\widehat{\vect{c}}_{\mathrm D}^{(\mathrm{RSS})}
=
\operatorname*{arg\,min}_{\vect{c}\in\mathcal{B}}
J_{\mathrm{RSS}}(\vect{c}),
\label{eq:rss_position_estimate}
\end{equation}
with the corresponding orientation estimate \(\widehat{\vect{n}}_{\mathrm D}=\widehat{\vect{n}}(\widehat{\vect{c}}_{\mathrm D}^{(\mathrm{RSS})})\) in the unknown-orientation case and \(\widehat{\vect{n}}_{\mathrm D}=\vect{n}_{\mathrm D}\) otherwise. In practice, \eqref{eq:rss_position_estimate} is evaluated over a discretisation of \(\mathcal{B}\), with the delay-based solution \(\widehat{\vect{c}}_{\mathrm D}^{(\mathrm{del})}\) serving both as the initialisation of the local refinement, \(\vect{c}_{0}^{(\mathrm{RSS})}=\widehat{\vect{c}}_{\mathrm D}^{(\mathrm{del})}\), and as an explicitly retained candidate through evaluation of \(J_{\mathrm{RSS}}(\widehat{\vect{c}}_{\mathrm D}^{(\mathrm{del})})\).
\begin{remark}
Delay-based position retention constitutes a safeguard because \eqref{eq:rss_position_estimate} minimises an amplitude-consistency criterion, therefore, its reliability is contingent on \(\widehat{\boldsymbol{\alpha}}\) being an unbiased representation of the \ac{LoS} gains. When this premise fails, for instance under unresolved \ac{NLoS} contamination, the \ac{RSS} objective may result in a misleading refinement direction. Retaining \(\widehat{\vect{c}}_{\mathrm D}^{(\mathrm{del})}\) in the candidate set guarantees that the \ac{RSS} estimator can never underperform the delay-only baseline. The composite estimator therefore inherits the robustness of delay-based geometry while retaining the sub-quantisation resolution of amplitude fitting.
\end{remark}
The \ac{RSS} branch estimates the receiver position by identifying the candidate whose predicted gain is most consistent with \(\widehat{\boldsymbol{\alpha}}\), and the link-wise ranges are subsequently obtained geometrically from the resulting position. For each usable link \(j\in\mathcal{J}_{\mathrm v}\), the \ac{RSS}-implied range is given as:
\begin{equation}
\widehat{d}_{0,j}^{(\mathrm{RSS})}
=\norm{
\widehat{\vect{c}}_{\mathrm D}^{(\mathrm{RSS})}
-\vect{c}_{\mathrm S,j}}.
\label{eq:rss_implied_range}
\end{equation}
Accordingly, \(\widehat{d}_{0,j}^{(\mathrm{RSS})}\) is the geometric distance implied by the estimated position, in contrast to \(\widehat{d}_{0,j}\), which is obtained from the detected \ac{LoS} tap prior to any multilateration. This distinction explains why \(\widehat{d}_{0,j}^{(\mathrm{RSS})}\) is not confined to multiples of \(\Delta d\) and can therefore achieve sub-bin ranging accuracy.

Reliable estimation requires a sufficient number of geometrically independent links. A localisation estimate is formed only when \(|\mathcal{J}_{\mathrm v}|\geq4\), which provides the redundancy required to identify the three positional degrees of freedom. For the unknown \ac{PD} orientation case, \(|\mathcal{J}_{\mathrm v}|\geq6\) is needed. Beyond link count, the spatial configuration must itself be informative, i.e., for unknown orientation, for solution to problem \eqref{eq:rss_candidate_normal_general}, \(
\operatorname{rank}\!\left(\vect{A}(\vect{c})\right)=3\) is a necessary condition. This condition requires the weighted incidence-direction vectors \(\{\kappa_{j}(\vect{c})\vect{u}_{\mathrm{DS},j}(\vect{c})\}_{j\in\mathcal{J}_{\mathrm v}}\) to span \(\mathbb{R}^3\), i.e., to be non-coplanar. Candidate positions for which \(\vect{A}(\vect{c})\) is rank deficient, severely ill-conditioned, or inconsistent with the imposed geometric constraints are therefore treated as infeasible.

\begin{remark}
For ceiling-mounted luminaires at a common height, the direction vectors \(\vect{u}_{\mathrm{DS},j}(\vect{c})\) become coplanar when the \acp{LED} are collinear in the ceiling plane, since the candidate position and any straight line of \acp{LED} jointly define a single plane. Orientation observability therefore depends on the \ac{LED} layout rather than on the \ac{SNR}. A collinear luminaire arrangement leaves the receiver normal partially unidentifiable regardless of how accurately \(\widehat{\boldsymbol{\alpha}}\) is estimated.
\end{remark}
\vspace{-4mm}
\subsection{Effect of Unresolved Multipath on LoS Gain Extraction}
In the \ac{LoS} scenario, the phase-compensated averaging operation 
in \eqref{eq:alpha_hat} coherently aligns the direct-path component across the occupied subcarriers, yielding \(\widehat{\alpha}_{0,j}\approx|\alpha_{0,j}|\). In a dispersive multipath channel, however, the estimated \ac{CFR} is a superposition of the \ac{LoS} component and \ac{NLoS} reflections, each characterised by a gain and delay tap. For the \ac{PF} subcarriers, the \ac{CFR} can therefore be expressed as:
\begin{equation}
H_{j,+}[n]
=
\sum_{\rho=0}^{P_j-1}
\alpha_{\rho,j}
e^{-\frac{\jmath2\pi n\lambda_{\rho,j}}{N}},
\label{eq:path_based_cfr_model}
\end{equation}
for \(n\in\mathcal{N}_{+}\), where \(\alpha_{\rho,j}\) denotes the gain of the \(\rho\)-th path and \(\lambda_{\rho,j}\) its associated discrete delay tap. Substituting \eqref{eq:path_based_cfr_model} into the gain-extraction operator of \eqref{eq:alpha_hat} yields:
\begin{equation}
\widehat{\alpha}_{0,j}
\approx
\left|
\alpha_{0,j}
+
\sum_{\rho=1}^{P_j-1}
\alpha_{\rho,j}
\Lambda_{\rho,j}
\right|,
\label{eq:alpha_hat_multipath_bias}
\end{equation}
where the leakage coefficient \(\Lambda_{\rho,j}=\frac{1}{|\mathcal{N}_{+}|}\sum_{n\in\mathcal{N}_{+}}e^{-\frac{\jmath2\pi n(\lambda_{\rho,j}-\widehat{\lambda}_{0,j})}{N}}\) quantifies the residual, uncompensated phase rotation retained by the \(\rho\)-th \ac{NLoS} component once phase compensation has been applied. \(\Lambda_{\rho,j}\) therefore contaminates the extracted \ac{LoS} gain estimate. Its magnitude is entirely governed by the tap separation \(\lambda_{\rho,j}-\widehat{\lambda}_{0,j}\) (depends on \(B\)), which dictates the rate at which the residual exponential term rotates across the subcarrier index \(n\). When the \(\rho\)-th path is sufficiently separated in the delay domain, i.e., its tap lies sufficiently far from \(\widehat{\lambda}_{0,j}\), the residual phasor sweeps through multiple full rotations as \(n\) traverses \(\mathcal{N}_{+}\), and the \ac{FD} averaging drives \(|\Lambda_{\rho,j}|\) toward zero through near-complete destructive interference. The reflected component is then suppressed from \(\widehat{\alpha}_{0,j}\), and \eqref{eq:alpha_hat_multipath_bias} reduces asymptotically to the ideal single-path result \(\widehat{\alpha}_{0,j}\approx|\alpha_{0,j}|\). Conversely, when an \ac{NLoS} component is strong, unresolved, or falls within or adjacent to the delay bin of the detected \ac{LoS} tap, i.e., \(|d_{\rho,j}-d_{0,j}|\lesssim\Delta d\), where \(d_{\rho,j} = c_0 \tau_{\rho,j}\), the phasor rotates too slowly for cancellation, and \(\Lambda_{\rho,j}\) attains a non-negligible magnitude, saturating at \(\Lambda_{\rho,j}=1\) in the worst case \(\lambda_{\rho,j}=\widehat{\lambda}_{0,j}\), where the reflection adds coherently rather than destructively to the gain estimate. Any such unresolved or closely spaced reflection therefore biases \(\widehat{\alpha}_{0,j}\), degrading \ac{RSS}-based localisation accuracy since the distorted gain propagates through the \ac{RSS} model into erroneous distance estimates, candidate positions, or incidence-angle geometry.
\begin{algorithm}[t]
\caption{Delay-Aided RSS-based Localisation}
\label{alg:rss_localisation_compact}
\begin{spacing}{1}
\begin{algorithmic}[1]
\REQUIRE
$\{\vect{c}_{\mathrm S,j},\vect{n}_{\mathrm S,j},
\widehat{\lambda}_{0,j},\widehat{\alpha}_{0,j}\}_{j=1}^{J}$, \(\widehat{\vect{c}}_{\mathrm D}^{(\mathrm{del})}\), $B$, search region $\mathcal{B}$,
and $\vect{n}_{\mathrm D}$ when the \ac{PD} orientation is known
\ENSURE
$\widehat{\vect{c}}_{\mathrm D}^{(\mathrm{RSS})}$,
$\{\widehat{d}_{0,j}^{(\mathrm{RSS})}\}_{j\in\mathcal{J}_{\mathrm v}}$,
and $\widehat{\vect{n}}_{\mathrm D}$ when required
\STATE Determine the usable \ac{LED} set $\mathcal{J}_{\mathrm v}$ and form
$\widehat{\boldsymbol{\alpha}}$.
\STATE Compute
$\widehat{d}_{\mathrm{tap},j}
=
\frac{c_0\widehat{\lambda}_{0,j}}{B}$
for all $j\in\mathcal{J}_{\mathrm v}$.
\STATE Obtain
$\widehat{\vect{c}}_{\mathrm D}^{(\mathrm{del})}$
from \eqref{eq:delay_multilateration} and use it to initialise the
\ac{RSS} search.
\FOR{each candidate position $\vect{c}\in\mathcal{B}$}
    \STATE Compute the candidate geometry, $\kappa_j(\vect{c})$, and
    construct $\vect{A}(\vect{c})$.
    \STATE Determine
    $\widehat{\vect{n}}(\vect{c})$
    using \eqref{eq:rss_candidate_normal_general}.
    \IF{$\mathcal{N}(\vect{c})=\varnothing$}
        \STATE Set $J_{\mathrm{RSS}}(\vect{c})=+\infty$.
    \ELSE
        \STATE Compute
        $\widehat{\boldsymbol{\alpha}}^{(\mathrm{pred})}(\vect{c})
        =
        \vect{A}(\vect{c})\widehat{\vect{n}}(\vect{c})$.
        \STATE Evaluate $J_{\mathrm{RSS}}(\vect{c})$ using
        \eqref{eq:rss_position_log_cost}.
    \ENDIF
\ENDFOR
\STATE Retain
$\widehat{\vect{c}}_{\mathrm D}^{(\mathrm{del})}$
as a fallback candidate.
\STATE Select
\(
\widehat{\vect{c}}_{\mathrm D}^{(\mathrm{RSS})}
=
\operatorname*{arg\,min}_{\vect{c}\in\mathcal{B}}
J_{\mathrm{RSS}}(\vect{c}).
\)
\STATE Set
$\widehat{\vect{n}}_{\mathrm D}
=
\widehat{\vect{n}}(
\widehat{\vect{c}}_{\mathrm D}^{(\mathrm{RSS})})$
when the orientation is unknown; otherwise,
$\widehat{\vect{n}}_{\mathrm D}=\vect{n}_{\mathrm D}$.
\STATE Compute
\(
\widehat{d}_{0,j}^{(\mathrm{RSS})}
=
\norm{
\widehat{\vect{c}}_{\mathrm D}^{(\mathrm{RSS})}
-
\vect{c}_{\mathrm S,j}
}\), for \(j\in\mathcal{J}_{\mathrm v}\).
\RETURN
$\widehat{\vect{c}}_{\mathrm D}^{(\mathrm{RSS})}$,
$\{\widehat{d}_{0,j}^{(\mathrm{RSS})}\}_{j\in\mathcal{J}_{\mathrm v}}$,
and $\widehat{\vect{n}}_{\mathrm D}$ when required.
\end{algorithmic}
\end{spacing}
\end{algorithm}
\vspace{-4mm}
\section{Performance Evaluation}
\label{sim_diss}
\begin{table}[t]
\caption{General Simulation Parameters}
\label{tab:simulation_parameters}
\centering
\footnotesize
\renewcommand{\arraystretch}{1.04}
\setlength{\tabcolsep}{1.8pt}
\begin{tabular}{
@{}
>{\raggedright\arraybackslash}p{3.30cm}
>{\centering\arraybackslash}p{1.25cm}
>{\raggedleft\arraybackslash}p{2.95cm}
@{}}
\toprule
\textbf{Parameter} & \textbf{Symbol} & \textbf{Value} \\
\midrule
\multicolumn{3}{@{}l}{\textbf{Indoor Optical Configuration}}\\
\midrule
Room dimensions
& {\fontsize{6.5pt}{7.5pt}\selectfont
   \(L_x\!\times\!L_y\!\times\!L_z\)}
& \(8\times8\times3.5~\si{\meter}\) \\
Number of \acp{LED}
& \(J\)
& \(6\) \\
\multirow{3}{*}{\ac{LED} coordinates}
& \(x_{\mathrm{LED}}\)
& \(\{1,4,7\}~\si{\meter}\) \\
& \(y_{\mathrm{LED}}\)
& \(\{2,6\}~\si{\meter}\) \\
& \(z_{\mathrm{LED}}\)
& \(3.5~\si{\meter}\) \\
\ac{PD} height bounds
& \(\{z_\mathrm{min},z_
\mathrm{max}\}\)
& \(\{0.2,1.5\}~\si{\meter}\) \\
Half-power semi-angle
& \(\phi_{1/2}\)
& \(60^\circ\) \\
Lambertian order
& \(q\)
& \(1\) \\
\ac{PD} active area
& \(A_{\mathrm{PD}}\)
& \(10^{-4}~\si{\meter\squared}\) \\
Optical-filter gain
& \(g_{\mathrm{of}}\)
& \(2.5481\) \\
Concentrator refractive index
& \(n_{\mathrm c}\)
& \(1.5\) \\
\ac{PD} \ac{FoV} semi-angle
& \(\Psi_{\mathrm{FoV}}\)
& \(60^\circ\) \\
Concentrator gain
& \(g_{\mathrm{oc}}\)
& \(3\) \\
\midrule
\multicolumn{3}{@{}l}{\textbf{Waveform and Simulation Parameters}}\\
\midrule
Number of subcarriers
& \(N\)
& \(\{64,256,1024,4096\}\) \\
Subcarrier spacing
& \(\Delta f\)
& \(240~\si{\kilo\hertz}\) \\
Sampling interval
& \(T_{\mathrm{sam}}\)
& \(\frac{1}{N\Delta f}\) \\
Number of \ac{OFDM} symbols
& \(N_{\mathrm{sym}}\)
& \(150\) \\
\ac{CP} Length
& \(N_{\mathrm{CP}}\)
& \(0.25N\) \\
Monte Carlo realisations
& \(N_{\mathrm{MC}}\)
& \(5000\) \\
Modulation order
& \(M_{\mathrm{QAM}}\)
& \(4\) \\
Candidate group sizes
& \(g_\ell\)
& \(\{16,8,4\}\) \\
Communication subcarriers
& \(k_\ell\)
& \(\{11,5,2\}\) \\
Sensing subcarriers
& \(p_\ell\)
& \(\{4,2,1\}\) \\
\ac{LoS} detection threshold
&
\(\varkappa_j\)
&
\(0.2\max_\lambda|\widehat{h}_{j,\mathrm{fr}}[\lambda]|\)\\
\midrule
\multicolumn{3}{@{}l}{\textbf{Energy Allocation Optimisation}}\\
\midrule
Reference communication energy
& \(\gamma_{\mathrm c}^{\mathrm{ref}}\)
& \(1\) \\
Reference sensing energy
& \(\gamma_{\mathrm s}^{\mathrm{ref}}\)
& \(1\) \\
Minimum sensing margin
& \(\Delta_{\mathrm s}\)
& \(3~\si{\decibel}\) \\
Feasible set for \(\mu\)
& \(\mathcal M\)
& \(\{0.51,0.53,\ldots,0.99\}\) \\
Communication weight
& \(w_{\mathrm c}\)
& \(1\) \\
Localisation weight
& \(w_{\mathrm s}\)
& \(1\) \\
\ac{BER} target
& \(F_{\mathrm c}^{\mathrm{tar}}\)
& \(10^{-3}\) \\
Position-\ac{RMSE} target
& \(F_{\mathrm s}^{\mathrm{tar}}\)
& \(0.5~\si{\meter}\) \\
Optimisation \acp{SNR}
& \(\mathcal S_{\mathrm{opt}}\)
& \(\{0,5,10,15,20\}~\si{\decibel}\) \\
\bottomrule
\end{tabular}
\end{table}

This section presents performance evaluation of \ac{SA}-\ac{OFDM}-\ac{IM} waveform in terms of both communication and sensing functionalities.  Simulation parameters are listed in Table~\ref{tab:simulation_parameters}. Unless otherwise stated, results reported hereafter are obtained using these default settings. \ac{DCO}-\ac{OFDM}, biased at \(12~\si{\decibel}\), is adopted as the communication benchmark, while the hybrid-\ac{RSS} estimator of \cite{xu2025integrated} serves as the localisation benchmark. Both waveforms employ \(4\)-\ac{QAM} (\(M_{\mathrm{QAM}}=4\)).

Unless otherwise stated, the results reported throughout this section are conditioned on full \ac{LED} visibility, i.e., all \(J=6\) \acp{LED} lie simultaneously within the \ac{PD}'s \ac{FoV}. 
This operating condition is retained for three reasons. First, both the delay-based multilateration stage and the proposed \ac{RSS}-based estimator solve an over-determined system constructed from the set of visible \ac{LED} links. Maximising the number of concurrent links, \(|\mathcal{J}_{\mathrm v}|=J\), yields the best-conditioned observation matrix \(\vect{A}(\vect{c})\), which is particularly consequential when the receiver orientation is estimated jointly with position. Evaluating performance under full visibility therefore isolates the best-case accuracy of the proposed waveform and localisation framework, independent of anchor-outage effects. Second, because the number of visible \acp{LED} is itself a random variable governed by the randomly drawn \ac{PD} position and orientation, allowing it to vary freely across Monte Carlo trials would confound the reported performance metrics with a variable-cardinality observation set, precluding a controlled, like-for-like comparison across \(N\), across different orientation scenarios, and against the \ac{DCO}-\ac{OFDM} and hybrid-\ac{RSS} benchmarks. 
Third, given the considered room dimensions, the six-\ac{LED} layout, and the \(60^\circ\) \ac{PD} \ac{FoV}, full visibility constitutes a representative, rather than artificial, operating regime for this deployment density. Extending the analysis to anchor outage conditions is left as a direction for future work.
\subsection{Energy Allocation Trade-off and Selected Operating Point}
\begin{figure}[!t]
    \centering
    \includegraphics[
        width=0.9\columnwidth,
        trim=0.2cm 0.2cm 0.18cm 0.2cm,
        clip
    ]{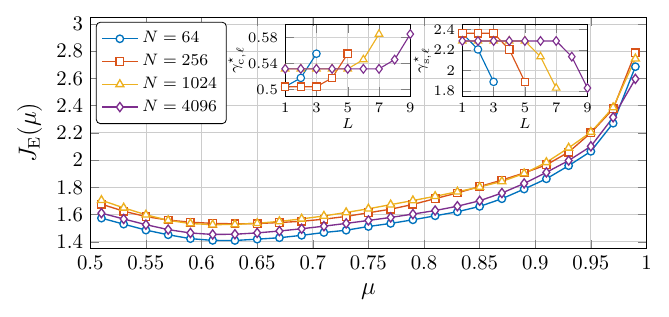}
    \caption{ \(J_{\mathrm E}(\mu)\) versus \(\mu\) for different
    \(N\) under a \ac{LoS} channel with an upward-facing \ac{PD}. The
    insets show the layer-dependent optimal energies
    \(\gamma^\star_{\mathrm c,\ell}\) and \(\gamma^\star_{\mathrm s,\ell}\) obtained using the corresponding common optimum \(\mu^\star\).}
    \label{fig:opt_upward}
\end{figure}

Fig.~\ref{fig:opt_upward} characterises the energy-allocation operating point of \ac{SA}-\ac{OFDM}-\ac{IM} under a \ac{LoS} channel with an upward-facing \ac{PD}, and sensing margin of \(\Delta_s\). 
As shown in Fig.~\ref{fig:opt_upward}, for the given configuration, \(J_{\mathrm E}(\mu)\) is minimised at \(\mu^\star\in[0.61,0.65]\), with the exact optimum depending on \(N\). Beyond \(\mu^\star\), additional sensing energy yields only marginal gains, since sensing subcarrier detection has already entered a high-reliability regime, while the allocated communication-subcarrier energy continues to decrease linearly with \(1-\mu\). At sufficiently large \(\mu\), the communication-tone energy approaches the null-tone noise floor, making the active communication subcarriers indistinguishable from inactive ones and degrading \ac{IM} pattern recovery. Consequently, \(\mu^\star\) marks the point at which the diminishing returns in sensing-pilot detection are exactly balanced against the rising average \ac{BER}. The resulting operating point is sensing-dominant, but not sensing-exclusive, since a non-negligible fraction of the reference energy is always retained for communication.

The insets of Fig.~\ref{fig:opt_upward} further report the layer-optimal energies \(\gamma^\star_{\mathrm c,\ell}\) and \(\gamma^\star_{\mathrm s,\ell}\) evaluated at \(\mu^\star\). Although \(\mu^\star\) is identical across layers, the resulting per-subcarrier energies vary with \(\ell\), because the communication and sensing subcarrier counts, \(k_\ell\) and \(p_\ell\), together with the layer-dependent reference energy \(E_\ell^{\mathrm{ref}}\). Layers sharing identical group parameters consequently yield identical energy values. Across the considered layer configurations, \(\gamma^\star_{\mathrm c,\ell}\) increases from approximately \(0.50\) to \(0.56\), while \(\gamma^\star_{\mathrm s,\ell}\) decreases from approximately \(2.36\) to \(1.89\). All layers nonetheless satisfy the minimum required \(\Delta_s\).
\vspace{-2mm}
\subsection{Communication Performance}
\subsubsection{Spectral Efficiency Analysis}
For layer \(\ell\), \(k_{\ell}\) active indices are selected from the \(g_{\ell}\) admissible intra-group positions, so that each group conveys \(\left\lfloor\log_{2}\binom{g_{\ell}}{k_{\ell}}\right\rfloor\) information bits through the index domain. Additionally, the \(k_{\ell}\) active communication subcarriers each carry a \(M_{\mathrm{QAM}}\)-ary \ac{QAM} symbol, contributing \(k_{\ell}\log_{2}M_{\mathrm{QAM}}\) bits per group. Since layer \(\ell\) comprises \(G_{\ell}\) groups, the aggregate bit payload for a given layer is:
\begin{equation}
\eta_{\ell} = G_{\ell}\left[\left\lfloor\log_{2}\binom{g_{\ell}}{k_{\ell}}\right\rfloor + k_{\ell}\log_{2}M_{\mathrm{QAM}}\right].
\label{eq:bits_layer}
\end{equation}
Summing \eqref{eq:bits_layer} over all \(L\) layers yields the total number of communication bits conveyed per \ac{SA}-\ac{OFDM}-\ac{IM} symbol, i.e.,
\(\eta_{\mathrm{OFDM}} = \sum_{\ell=1}^{L} \eta_\ell\). The corresponding \ac{SE}, expressed in \(\si{\bit\per\second\per\hertz}\), is obtained by normalising \(\eta_{\mathrm{OFDM}}\) by the total symbol duration in samples, including the \ac{CP} as:
\begin{equation}
\mathrm{SE} = \frac{\eta_{\mathrm{OFDM}}}{N + N_{\mathrm{CP}}}.
\label{eq:spectral_efficiency}
\end{equation}
Evaluating \eqref{eq:spectral_efficiency} excluding \ac{CP} overhead so as to isolate the modulation-domain efficiency of the waveform itself, yields \ac{SE} of \(0.8549\), \(1.0117\), \(1.0498\), and \(1.0562~\si{\bit\per\second\per\hertz}\) for \(N=64\), \(256\), \(1024\), and \(4096\), respectively, for \ac{SA}-\ac{OFDM}-\ac{IM} while \ac{DCO}-\ac{OFDM} achieves \(0.9375\), \(0.9844\), \(0.9902\), and \(0.9912~\si{\bit\per\second\per\hertz}\) over the same range of \(N\). This implies that \ac{SA}-\ac{OFDM}-\ac{IM} is inherently more spectral efficient compared to \ac{DCO}-\ac{OFDM}.
\vspace{-3mm}
\subsubsection{Bit-Error Rate Performance}
\begin{figure}[!t]
    \centering
    \includegraphics[
        width=0.9\columnwidth,
        trim=0.2cm 0.2cm 0.18cm 0.2cm,
        clip
    ]{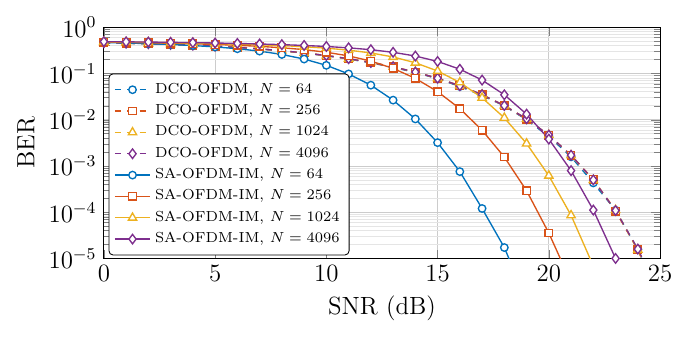}
    \caption{\ac{BER} performance of \ac{SA}-\ac{OFDM}-\ac{IM} and \ac{DCO}-\ac{OFDM} versus \ac{SNR} under \ac{LoS} propagation with an upward-facing \ac{PD} for different \(N\).}
    \label{fig:ber_upward}
\end{figure}

Fig.~\ref{fig:ber_upward} benchmarks the \ac{BER} performance of \ac{SA}-\ac{OFDM}-\ac{IM} against \ac{DCO}-\ac{OFDM} under \ac{LoS} propagation with an upward-facing \ac{PD}. The reported \ac{BER} is obtained by first evaluating the link-level \ac{BER} independently for each \ac{LED}-to-\ac{PD} pair and subsequently averaging across all \acp{LED}. It can be seen that \ac{SA}-\ac{OFDM}-\ac{IM} consistently outperforms \ac{DCO}-\ac{OFDM} across all considered \(N\). At a target \ac{BER} of \(10^{-3}\), \ac{SA}-\ac{OFDM}-\ac{IM} requires approximately \(16\), \(18.5\), \(20.3\), and \(21.3~\si{\decibel}\) for \(N=64\), \(256\), \(1024\), and \(4096\), respectively, whereas \ac{DCO}-\ac{OFDM}, implemented following the framework of \cite{xu2025integrated} with the first \ac{OFDM} symbol of each frame reserved for \ac{CE}, requires approximately \(21\)-\(22.5~\si{\decibel}\). The largest \ac{SNR} advantage of the proposed scheme is observed at \(N=64\), where a gain of approximately \(5~\si{\decibel}\) is achieved relative to \ac{DCO}-\ac{OFDM}. This \ac{BER} advantage is obtained without an \ac{SE} penalty. \ac{DCO}-\ac{OFDM} attains a marginally higher \ac{SE} only at \(N=64\), whereas \ac{SA}-\ac{OFDM}-\ac{IM} achieves both a higher \ac{SE} and a lower \ac{BER} for \(N\geq256\). Furthermore, the \ac{BER} of \ac{SA}-\ac{OFDM}-\ac{IM} increases with \(N\), an effect primarily attributable to the larger index-domain search space at higher \(N\). A greater number of subcarriers increases the likelihood of erroneous \ac{IM} subcarriers detection, and each such detection error subsequently causes incorrect extraction and demodulation of the associated \ac{QAM} symbols, compounding into a higher overall \ac{BER}.
\subsubsection{Channel Estimation Accuracy}
\begin{figure}[!t]
    \centering
    \includegraphics[
        width=0.85\columnwidth,
        trim=0.2cm 0.2cm 0.18cm 0.1cm,
        clip
    ]{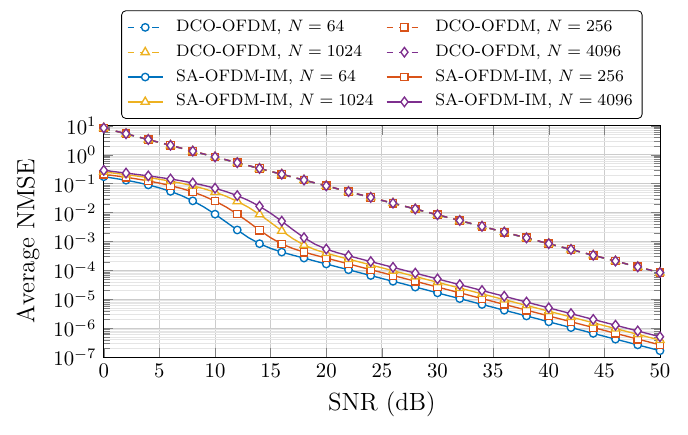}
   \caption{Average \ac{CE} \ac{NMSE} of
\ac{SA}-\ac{OFDM}-\ac{IM} and \ac{DCO}-\ac{OFDM} versus \ac{SNR} under \ac{LoS} propagation with an upward-facing \ac{PD} for different \(N\).}
    \label{fig:nmse_upward}
\end{figure}

The \ac{CE} accuracy is quantified through the average \ac{NMSE}, defined as:
\begin{equation}
\mathrm{NMSE} = \frac{1}{N_{\mathrm{MC}}|\mathcal{J}_{\mathrm v}|}\sum_{\zeta=1}^{N_{\mathrm{MC}}}\sum_{j\in\mathcal{J}_{\mathrm v}} \frac{\norm{\widehat{\vect{h}}_{j,\mathrm{fr}}^{(\zeta)} - \vect{h}_{j}^{(\zeta)}}^2}{\norm{\vect{h}_{j}^{(\zeta)}}^2},
\end{equation}
where \(\vect{h}_{j}^{(\zeta)}\) and \(\widehat{\vect{h}}_{j,\mathrm{fr}}^{(\zeta)}\) denote the true and estimated \ac{CIR} vectors, respectively, for the \(j\)-th \ac{LED} link in the \(\zeta\)-th Monte Carlo realisation. 
Fig.~\ref{fig:nmse_upward} compares the \ac{CE} \ac{NMSE} of \ac{SA}-\ac{OFDM}-\ac{IM} and \ac{DCO}-\ac{OFDM} under \ac{LoS} propagation with an upward-facing \ac{PD}. From Fig.~\ref{fig:nmse_upward}, we observe that \ac{SA}-\ac{OFDM}-\ac{IM} achieves a lower \ac{NMSE} than \ac{DCO}-\ac{OFDM} because of reduced noise variance obtained by averaging the pilot-based \ac{CFR} estimates across repeated observations, which are repeated across multiple OFDM symbols. In contrast, \ac{DCO}-\ac{OFDM} relies exclusively on a single dedicated training symbol for \ac{CE}, yielding a substantially smaller observation set.

At low \acp{SNR}, the \ac{NMSE} of \ac{SA}-\ac{OFDM}-\ac{IM} is dominated by sensing-pilot detection errors that contaminate the resulting \ac{CFR} estimate. As \ac{SNR} increases, the sensing-to-communication energy separation improves subcarrier detectability, producing the pronounced \ac{NMSE} reduction seen over the intermediate-\ac{SNR} regime. At high \acp{SNR}, the residual \ac{NMSE} is dominated by interpolation distortion across the unobserved subcarriers. The \ac{NMSE} rises marginally with \(N\) since a larger \(N\) increases the number of channel coefficients requiring interpolation between sensing pilots. \ac{SA}-\ac{OFDM}-\ac{IM} nevertheless retains a clear advantage over \ac{DCO}-\ac{OFDM} across all considered \(N\). At \(50~\si{\decibel}\), it achieves an average \ac{NMSE} of the order of \(10^{-7}\)-\(10^{-6}\) against \(10^{-4}\) for \ac{DCO}-\ac{OFDM}.
%
\subsection{Sensing Performance Analysis}
Table~\ref{tab:orientation_scenarios} summarises the three \ac{PD} orientation models used to evaluate localisation performance. The upward-facing case fixes \(\theta_{\mathrm D}=0\) with a known, static normal vector \(\vect{n}_{\mathrm D}\). For the random-orientation cases, the polar tilt is modelled as Laplace-distributed, \(\theta_{\mathrm D}\sim\mathcal{L}(\mu_\theta,b_\theta)\) with \(\mu_\theta=29.67^\circ\) and \(b_\theta=7.78^\circ\), consistent with the empirically validated device-tilt statistics of \cite{soltani2018modeling}, while the azimuthal orientation is drawn uniformly, \(\omega_{\mathrm D}\sim\mathcal{U}[-180^\circ,180^\circ]\). The instantaneous normal \(\vect{n}_{\mathrm D}\) is treated as either known or unknown to the estimator, yielding the two remaining scenarios in Table~\ref{tab:orientation_scenarios}. In all three cases, the \ac{PD} Cartesian coordinates are drawn uniformly at random within the room dimensions.

%
\begin{table}[t]
\caption{Receiver Orientation Scenarios}
\label{tab:orientation_scenarios}
\centering
\footnotesize
\renewcommand{\arraystretch}{1.12}
\setlength{\tabcolsep}{0.6pt}
\begin{tabular}{
@{}
>{\raggedright\arraybackslash}p{2.20cm}
>{\centering\arraybackslash}p{1.65cm}
>{\centering\arraybackslash}p{2.2cm}
>{\centering\arraybackslash}p{1.75cm}
@{}}
\toprule
\textbf{\ac{PD} Orientation}
& \({\theta_{\mathrm D}}\)
& \({\omega_{\mathrm D}}\)
& \textbf{\ac{PD} knowledge} \\
\midrule
Upward-facing
& \(0\)
& --
& \(\vect n_{\mathrm D}\) known and fixed \\
Random, known
& \(\mathcal{L}(\mu_\theta,b_\theta)\)
& \(\mathcal{U}[-180^\circ,180^\circ]\)
& \(\vect n_{\mathrm D}\) known \\
Random, unknown
& \(\mathcal{L}(\mu_\theta,b_\theta)\)
& \(\mathcal{U}[-180^\circ,180^\circ]\)
& \(\vect n_{\mathrm D}\) unknown \\
\bottomrule
\end{tabular}
\end{table}
\begin{figure}[!t]
    \centering
    \includegraphics[
        width=0.9\columnwidth,
        trim=0.2cm 0.2cm 0.18cm 0.1cm,
        clip
    ]{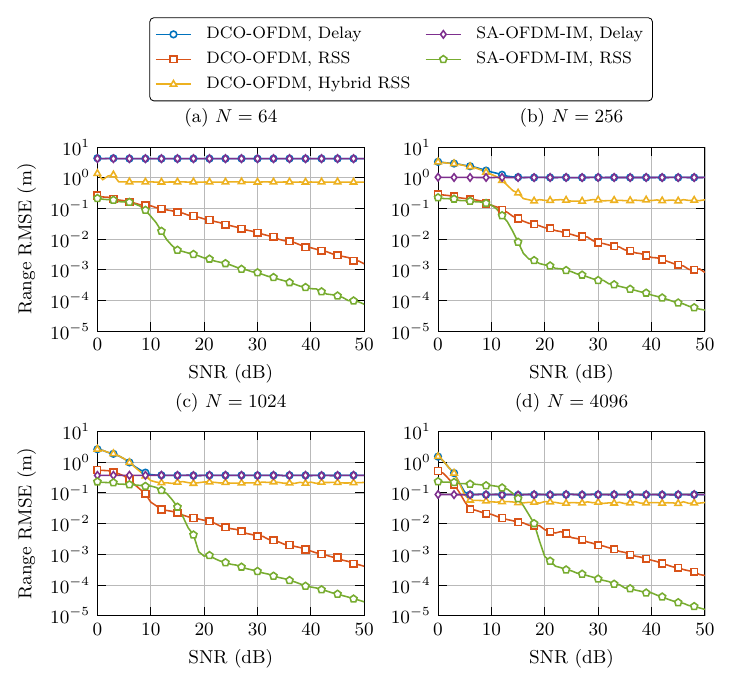}
  \caption{Range \ac{RMSE} in metres under \ac{LoS} propagation with an upward-facing \ac{PD} for different \(N\).}
    \label{fig:rmse_range_upward}
\end{figure}
\subsubsection{Upward-Facing \ac{PD} at Random Receiver Positions}
\begin{figure}[!t]
    \centering
    \includegraphics[
        width=0.9\columnwidth,
        trim=0.2cm 0.2cm 0.18cm 0.1cm,
        clip
    ]{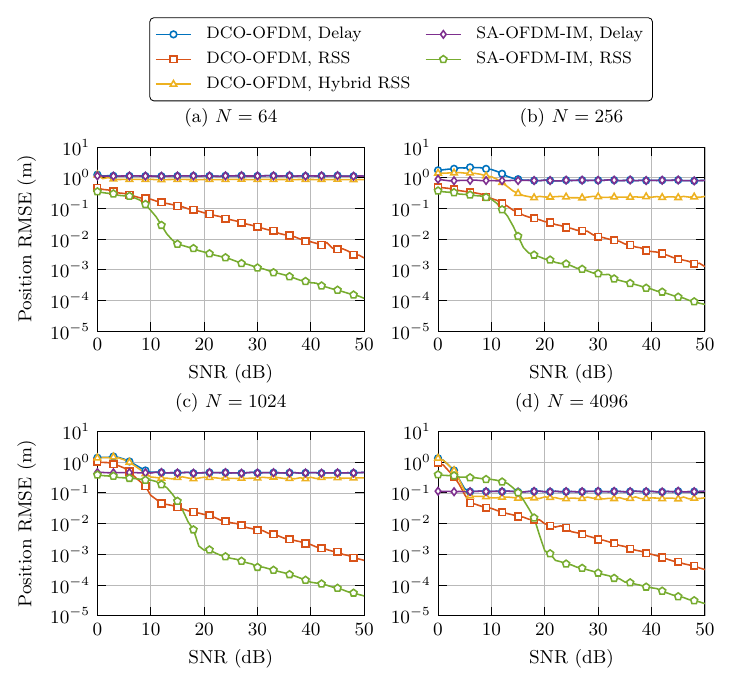}
    \caption{Position \ac{RMSE} in metres under \ac{LoS} propagation with an upward-facing \ac{PD} for different \(N\).}
    \label{fig:rmse_pos_upward}
\end{figure}

Figs.~\ref{fig:rmse_range_upward} and~\ref{fig:rmse_pos_upward} compare the range- and position-estimation performance achieved by the proposed framework against the benchmark in \cite{xu2025integrated}, for an upward-facing \ac{PD} and different \(N\). The range \ac{RMSE} is evaluated as \(\sqrt{\frac{1}{N_{\mathrm{MC}}|\mathcal{J}_{\mathrm v}|}\sum_{\zeta=1}^{N_{\mathrm{MC}}}\sum_{j\in\mathcal{J}_{\mathrm v}}\left(\widehat d_j^{(\zeta)}-d_{0,j}^{(\zeta)}\right)^2}\), whereas the position \ac{RMSE} is given as \(\sqrt{\frac{1}{N_{\mathrm{MC}}}\sum_{\zeta=1}^{N_{\mathrm{MC}}}\norm{\widehat{\vect{c}}_{\mathrm D}^{(\zeta)}-\vect{c}_{\mathrm D}^{(\zeta)}}^2}\).  \(\mu^\star\in[0.61,0.65]\) across the considered \(N\) for this scenario.

The delay-based estimators exhibit \ac{SNR}-independent error floors, since \(\widehat d_{\mathrm{tap},j}\) is restricted to integer multiples of the delay-bin resolution \(\Delta d\). This range-quantisation error propagates into the multilateration stage and consequently bounds the achievable position accuracy. Increasing \(N\) reduces these floors by refining \(\Delta d\). The proposed \ac{RSS}-based estimator, in contrast, achieves the lowest range and position \acp{RMSE} across all considered \(N\). For \ac{SA}-\ac{OFDM}-\ac{IM}, this advantage stems from the distributed sensing-pilot placement and frame-level averaging, which jointly improves the \ac{CFR} and \ac{LoS} gain estimates. The waterfall behaviour exhibited by \ac{RSS}-based localisation at low-to-moderate \acp{SNR} reflects residual sensing-pilot detection errors. Once reliable subcarrier recovery is achieved, both the range and position \acp{RMSE} decrease rapidly. Applying the same \ac{RSS} estimator to \ac{DCO}-\ac{OFDM} likewise improves its accuracy relative to delay-only estimation, but remains less precise than for \ac{SA}-\ac{OFDM}-\ac{IM}, since its \ac{CE} relies solely on a single dedicated training symbol. We further observe that the hybrid-\ac{RSS} method of \cite{xu2025integrated} improves upon delay-only estimation, but its accuracy remains fundamentally constrained by \(\Delta d\) and degrades further under randomly distributed receiver positions.
\subsubsection{Randomly Oriented \ac{PD} at Random Receiver Positions}
\begin{figure}[!t]
    \centering
    \includegraphics[
        width=0.9\columnwidth,
        trim=0.2cm 0.2cm 0.18cm 0.1cm,
        clip
    ]{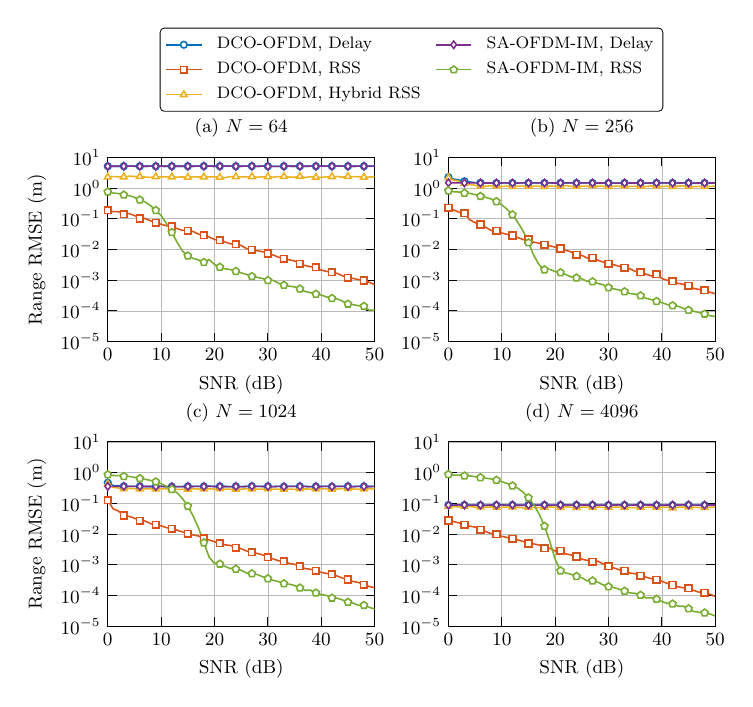}
    \caption{Range \ac{RMSE} in metres  under \ac{LoS} propagation with a random known \ac{PD} orientation for different \(N\).}
    \label{fig:rmse_range_known_6}
\end{figure}

\begin{figure}[!t]
    \centering
    \includegraphics[
        width=0.9\columnwidth,
        trim=0.2cm 0.2cm 0.18cm 0.1cm,
        clip
    ]{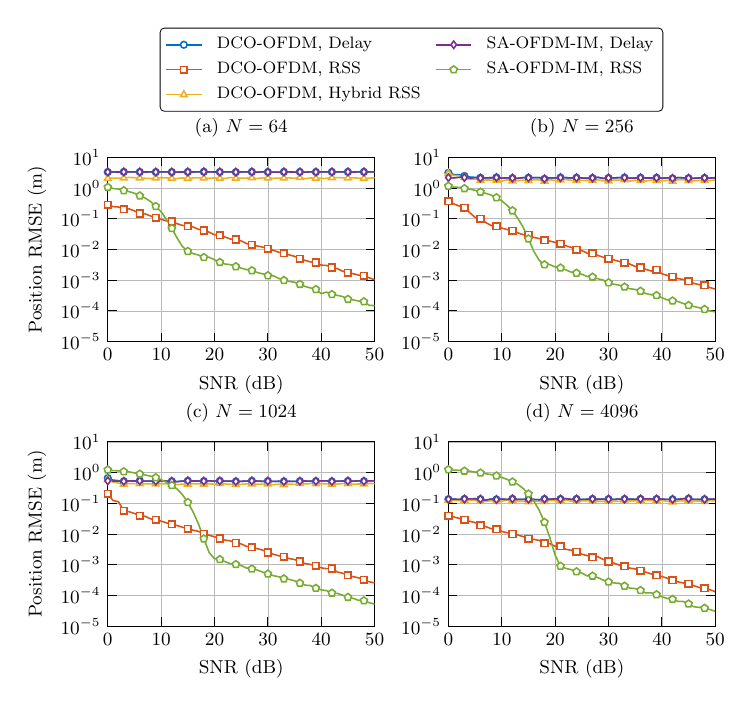}
    \caption{Position \ac{RMSE} in metres under \ac{LoS} propagation with a random known \ac{PD} orientation for different \(N\).}
    \label{fig:rmse_pos_known_6}
\end{figure}

Figs.~\ref{fig:rmse_range_known_6} and~\ref{fig:rmse_pos_known_6} report the range- and position-estimation performance for randomly oriented \ac{PD} whose instantaneous orientation is known to the estimator. The observed trends are consistent with those obtained for known \ac{PD} orientation case. Here, the optimised sensing-energy fraction is \(\mu^\star=0.63\) for \(N\in\{64,256\}\) and \(\mu^\star=0.65\) for \(N\in\{1024,4096\}\).

The proposed \ac{RSS}-based estimator again achieves the lowest range and position \acp{RMSE}. Although the random orientation alters the incidence angles and thereby introduces link-to-link variation in the received \ac{LoS} gains, this does not degrade estimation accuracy, since the known \ac{PD}-normal is incorporated explicitly into the predicted gain model. The predicted gain \(\alpha_{j}^{(\mathrm{pred})}(\vect{c},\vect{n}_{\mathrm D})\) accounts for the orientation-dependent incidence projection \(\vect{u}_{\mathrm{DS},j}^{\mathrm T}(\vect{c})\vect{n}_{\mathrm D}\). The \ac{RSS} cost function therefore compares the measured gains against orientation-consistent model predictions, preventing angular attenuation from being misattributed to a distance or position error. We further observe that the proposed \ac{RSS}-based estimator performs reasonably well when applied to \ac{DCO}-\ac{OFDM}, indicating that the estimator generalises beyond the proposed waveform to other optical \ac{OFDM} schemes. We also observe that the performance of hybrid-\ac{RSS} is severely impacted by unknown \ac{PD} orientation.
\subsubsection{Unknown \ac{PD} Orientation at Random Receiver Positions}
\begin{figure}[!t]
    \centering
    \includegraphics[
        width=0.9\columnwidth,
        trim=0.2cm 0.2cm 0.18cm 0.1cm,
        clip
    ]{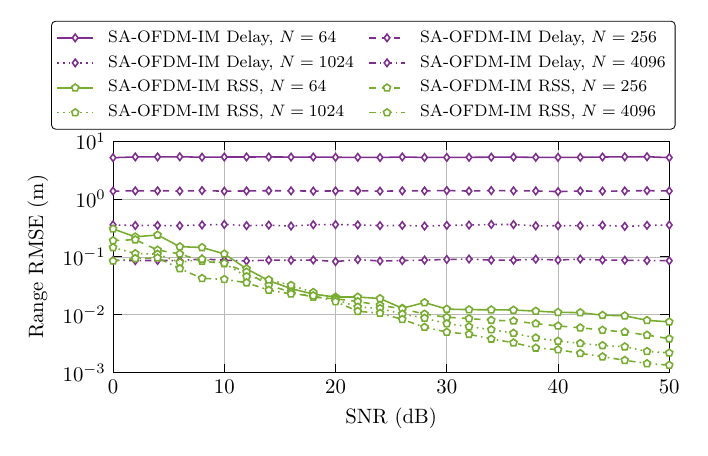}
    \caption{Range \ac{RMSE} in metres  under \ac{LoS} propagation with a random unknown \ac{PD} orientation for different \(N\).}
    \label{fig:rmse_range_unknown_6}
\end{figure}

\begin{figure}[!t]
    \centering
    \includegraphics[
        width=0.9\columnwidth,
        trim=0.2cm 0.2cm 0.18cm 0.1cm,
        clip
    ]{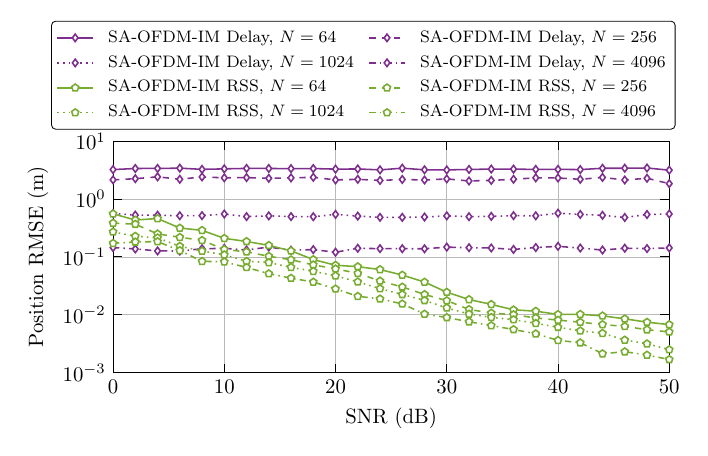}
    \caption{Position \ac{RMSE} in metres under \ac{LoS} propagation with a random unknown \ac{PD} orientation for different \(N\).}
    \label{fig:rmse_pos_unknown_6}
\end{figure}

\begin{figure}[!t]
    \centering
    \includegraphics[
        width=0.9\columnwidth,
        trim=0.2cm 0.2cm 0.18cm 0.1cm,
        clip
    ]{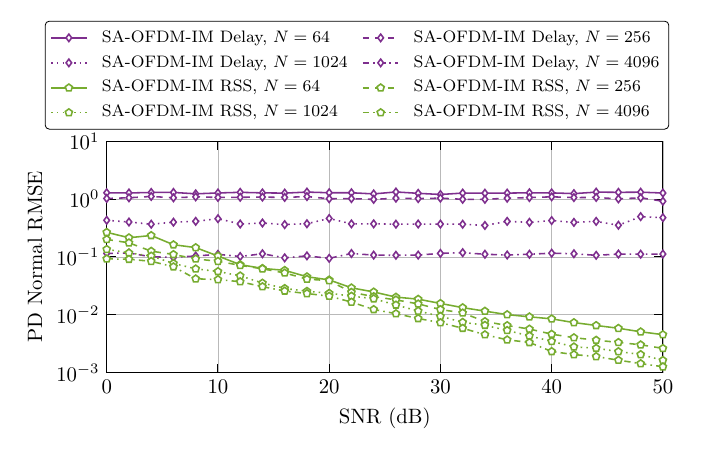}
    \caption{Orientation \ac{RMSE} under \ac{LoS} propagation with a random unknown \ac{PD} orientation for different \(N\).}
    \label{fig:rmse_orient_unknown_6}
\end{figure}

Figs.~\ref{fig:rmse_range_unknown_6}, \ref{fig:rmse_pos_unknown_6}, and~\ref{fig:rmse_orient_unknown_6} present the range, position, and \ac{PD}-normal estimation performance when the \ac{PD} orientation is random and unknown to the estimator. For each candidate position \(\vect{c}\), the corresponding normal is evaluated as \(\widehat{\vect{n}}(\vect{c})\), and the resulting predicted gain vector is used to assess the \ac{RSS} consistency of that candidate. The \ac{PD}-normal \ac{RMSE} is evaluated as \(\sqrt{\frac{1}{N_{\mathrm{MC}}}\sum_{\zeta=1}^{N_{\mathrm{MC}}}\norm{\widehat{\vect{n}}_{\mathrm D}^{(\zeta)}-\vect{n}_{\mathrm D}^{(\zeta)}}^2}\), with \(\mu^\star=0.61\) for \(N\in\{64,256,1024\}\) and \(\mu^\star=0.63\) for \(N=4096\).

The delay-based estimator again exhibits \ac{SNR}-independent error floors due to the delay-bin resolution \(\Delta d\). This quantisation error propagates through the multilateration stage and consequently degrades the associated orientation estimate, since the observation matrix \(\vect{A}(\widehat{\vect{c}}_{\mathrm D}^{(\mathrm{del})})\) is constructed from the quantised, delay-based position estimate. The \ac{RSS}-based estimator, in contrast, jointly exploits the measured gain pattern to refine both position and orientation. Its range and position \acp{RMSE} decrease continuously with \ac{SNR} and remain substantially below those of the delay-based method, and an analogous trend is observed for the \ac{PD}-normal \ac{RMSE}. This behaviour follows from the spatial diversity afforded by the multiple \ac{LED} links, each providing a distinct incidence direction. This diversity improves the conditioning of \(\vect{A}(\vect{c})\) and enables the unknown receiver normal to be recovered jointly with position. Relative to the upward-facing and known-orientation scenarios, however, \ac{RSS}-based localisation accuracy degrades once the \ac{PD} orientation becomes unknown, since the estimated normal vector directly enters the incidence-angle term of the predicted gain model. Any residual error in the recovered normal therefore induces a mismatch between the measured and predicted \ac{RSS} values, which can bias the position estimate and inflate the overall localisation error.
\vspace{-4mm}
\section{Conclusion}
\label{sec:conclusion}
This paper proposed \ac{SA}-\ac{OFDM}-\ac{IM} waveform for joint optical wireless communication and localisation. The design embeds communication symbols and sensing pilots across layered, index-modulated subcarrier groups, while an adaptive periodic bias enforces the non-negativity constraint. At the receiver, a low-complexity energy detector distinguishes communication from sensing subcarriers, and frame-level aggregation and averaging of the distributed sensing pilots yield robust \ac{CFR} and \ac{CIR} estimates, enabling reliable extraction of both the \ac{LoS} delay and its phase-compensated \ac{LoS} gain. The delay branch provides localisation constrained by the fundamental delay resolution of the system bandwidth, the proposed delay-aided \ac{RSS} estimator achieves sub-delay-bin accuracy. Moreover, when the \ac{PD} orientation is unknown, its normal vector can be estimated jointly with position through the candidate-dependent gain model. The reported results demonstrate that \ac{SA}-\ac{OFDM}-\ac{IM} outperforms \ac{DCO}-\ac{OFDM} in both \ac{BER} and \ac{CE} accuracy, while the proposed \ac{RSS}-based estimator consistently achieves the lowest range and position \acp{RMSE} among all considered benchmarks. These gains are attributable to distributed pilot embedding and the frame-level aggregation of repeated sensing subcarriers. The analysis further reveals that unresolved \ac{NLoS} components can bias the extracted \ac{LoS} gain. The development of dedicated multipath-suppression mechanisms is therefore left as a direction for future work, alongside the extension of the anchor-blockage conditions.
\vspace{-4mm}
\section*{Acknowledgment}
This work was supported by EPSRC project UKRI916 PerceptiFi.
\vspace{-5mm}
\bibliographystyle{IEEEtran}
\bibliography{biblio}
\end{document}